\RequirePackage{fix-cm}
\documentclass[smallextended]{svjour3}       
\smartqed  

\usepackage{afterpage}

\usepackage{url}
\usepackage[hidelinks]{hyperref}
\usepackage[utf8]{inputenc}
\usepackage[small]{caption}
\usepackage{graphicx}
\usepackage{booktabs}
\newcommand{\dquotes}[1]{``#1''}

\usepackage{xcolor}
\usepackage{verbatim}
\usepackage{multirow}

\usepackage{tabularx}
\usepackage{enumitem}
\usepackage{longtable}

\setlist{nolistsep}
\definecolor{green}{HTML}{66FF66}
\definecolor{myGreen}{HTML}{009900}

\definecolor{blue(pigment)}{rgb}{0.2, 0.2, 0.6}

\definecolor{cadmiumgreen}{rgb}{0.0, 0.42, 0.24}
\definecolor{djx}{rgb}{0.0, 0.5, 0.0}

\definecolor{ydx}{rgb}{0.8, 0.2, 0.1}

\definecolor{rev}{rgb}{0.0, 0.25, 0.95}

\def\review#1{{\color{black}#1}}
\usepackage[sort&compress]{natbib} 
\usepackage{todonotes}
\presetkeys
    {todonotes}
    {inline,backgroundcolor=yellow}{}

\journalname{UMUAI}

\begin{document}

\title{Fairness in Recommender Systems: Research Landscape and Future Directions}

\titlerunning{Fairness in Recommender Systems: Research Landscape and Future Directions}

\author{Yashar Deldjoo \and  Dietmar Jannach \and Alejandro Bellogin \and Alessandro Difonzo \and Dario Zanzonelli}

\authorrunning{Deldjoo et al.}

\institute{
    Y. Deldyoo, A. Difonzo, D. Zanzonelli \at
             Polytechnic University of Bari, Italy \\
              \email{deldjooy@acm.org}
           \and
           D. Jannach\at
              University of Klagenfurt, Austria \\
              \email{dietmar.jannach@aau.at}
           \and
           A. Bellogin \at
              University Autonomous of Madrid, Spain \\
              \email{alejandro.bellogin@uam.es}}
\date{April 24, 2023}
\maketitle

\begin{abstract}
Recommender systems can strongly influence which information we see online, e.g., on social media, and thus impact our beliefs, decisions, and actions. At the same time, these systems can create substantial business value for different stakeholders. Given the growing potential impact of such AI-based systems on individuals, organizations, and society, questions of \emph{fairness} have gained increased attention in recent years. However, research on fairness in recommender systems is still a developing area. In this survey, we first review the fundamental concepts and notions of fairness that were put forward in the area in the recent past.
Afterward, through a review of more than 160 scholarly publications, we present an overview of how research in this field is currently operationalized, e.g., in terms of general research methodology, fairness measures, and algorithmic approaches.
Overall, our analysis of recent works points to certain research gaps. In particular, we find that in many research works in computer science, very abstract problem operationalizations are prevalent and
\review{questions of the underlying normative claims and what represents a fair recommendation in the context of a given application are often not discussed in depth. These observations call for more interdisciplinary research to address fairness in recommendation in a more comprehensive and impactful manner.}
\end{abstract}

\section{Introduction}
Recommender systems (RS) are one of the most visible and successful applications of AI technology in practice, and personalized recommendations---as provided on many modern e-commerce or media sites--can have a substantial impact on different stakeholders. On e-commerce sites, for example, the choices of consumers can be largely influenced by recommendations, and these choices are often directly related to the profitability of the platform. On news websites or social media, on the other hand, personalized recommendations may determine to a large extent which information we see, which in turn may shape not only our own beliefs, decisions, and actions but also the beliefs of a community of users or an entire society.

In academia, recommenders have historically been considered as ``benevolent'' systems that create value for consumers, e.g., by helping them find relevant items, and that this value for consumers then translates to value for businesses, e.g., due to higher sales numbers or increased customer retention~\citep{jannachjugovactmis2019}. Only in the most recent years was more awareness raised regarding possible negative effects of automated recommendations, e.g., that they may promote items on an e-commerce site that mainly maximize the profit of providers or that they may lead to an increased spread of misinformation on social media.

Given the potentially significant effects of recommendations on different stakeholders, researchers increasingly argue that providing recommendations may raise various ethical questions and should thus be done in a \emph{responsible} way~\citep{DBLP:journals/widm/NtoutsiFGINVRTP20,trattner2022aiethics}. One important ethical question in this context is that of the \emph{fairness} of a recommender system, see~\citep{Burke2017Fairness,ekstrand2021fairness}, reflecting related discussions on the more general level of~\emph{fair machine learning} and~\emph{fair AI}~\citep{mehrabi2021fairnesssurvey,barocas-hardt-narayanan,DBLP:journals/widm/NtoutsiFGINVRTP20}.

During the last few years, researchers have discussed and analyzed different dimensions in which a recommender system should be fair or vice versa.

Given the nature of fairness as a social construct, it, however, seems difficult or even impossible~\citep{ekstrand2021fairness} to establish a general definition of what represents a fair recommendation. In addition to the subjectivity of fairness, there are frequently competing stakeholder interests to account for in real-world recommendation contexts~\citep{naghiaei2022cpfair,abdollahpouri2020}.

With this survey, we aim to provide an overview of what has been achieved in this emerging area so far and highlight potential research gaps. Specifically, drawing on an analysis of \review{more than 150} recent papers in computer science, we investigate \emph{(i)} which dimensions and definitions of fairness in RS have been identified and established, \emph{(ii)} at which application scenarios researchers target and which examples they provide, and \emph{(iii)} how they operationalize the research problem in terms of methodology, algorithms, and metrics. Based on this analysis, we then paint a landscape of current research in various dimensions and discuss potential shortcomings and future directions for research in this area.

Overall, we find that research in computing typically assumes that a clear definition of fairness is available, thus rendering the problem as one of designing algorithms to optimize a given metric. Such an approach may however appear too abstract and simplistic, cf.~\cite{selbst2019}, calling for more faceted and multi-disciplinary approaches to research in fairness-aware recommendation.

\review{The paper is organized as follows. Next, in Section~\ref{sec:background}, we lay out the motivation behind this survey in more detail, and we present the essential notions used to characterize fairness in the literature. Section~\ref{sec:methodology} then presents our methodology to identify and categorize relevant research works. Section~\ref{sec:landscape} represents the main part of our study, which paints the current research landscape of fairness in recommender systems in various dimensions, e.g., in terms of the addressed fairness problem and the chosen research methodology. In Section~\ref{sec:discussion}, we then reflect on these observations and identify open challenges and possible future research directions.}

\section{Background and \review{Foundations}}
\label{sec:background}

\subsection{Examples of Unfair Recommendations}

In the general literature on Fair ML/AI, an important application case is the automated prediction of recidivism by convicted criminal. In this case, an ML-based system is usually considered unfair if its predictions depend on demographic aspects like ethnicity and when it then ultimately discriminates members of certain ethnic groups~\review{\citep{angwin2016machine}}.
In the context of our present work, such use cases of ML-based decision-support systems are not in focus. Instead, we focus on common application areas of RS where~\emph{personalized} item suggestions are made to users, e.g., in e-commerce, media streaming, or news and social media sites.

At first sight, one might think that the recommendation providers here are independent businesses and it is entirely at their discretion which shopping items, movies, jobs, or social connections they recommend on their platforms. Also, one might assume that the~\emph{harm} that is made by such recommendations is limited, compared, e.g., to the legal decision problem mentioned above. There are, however, several situations also in typical application scenarios of RS where many people might think a system is unfair in some sense. For example, an e-commerce platform might be considered unfair if it mainly promotes those shopping items that maximize its own profit but not consumer utility. Besides such intentional interventions, there might also be situations where an RS reinforces existing discrimination patterns or biases in the data, e.g., when a system on an employment platform mainly recommends lower-paid jobs to certain demographic groups.

Nonetheless, questions of fairness in RS extend beyond the consumer's perspective. In reality, a recommendation service often involves multiple stakeholders~\citep{abdollahpouri2020}.
On music streaming platforms, for example, we have not only the consumers but also the artists, record labels, and the platform itself, which might have diverging goals that may be affected by the recommendation service. Artists and labels are usually interested in increasing their visibility through recommendations. On the other hand, platform providers might seek to maximize engagement with the service across the entire user base, which might result in promoting mostly already popular artists and tracks with the recommendations. Such a strategy, however, easily leads to a ``rich-get-richer'' effect and reduces the chances of less popular artists being exposed to consumers, which might be considered~\emph{unfair to providers}. Finally, there are also use cases where recommendations may have~\emph{societal} impact, particularly on news and social media sites. Some may consider it unfair if a recommender system only promotes content that emphasizes one side of a political discussion or promotes misinformation that is suitable to discriminate against certain user groups.

\review{As we will see later, different notions of fairness exist in the literature. What is important, however, is that in any discussed scenario, there are certain ethical questions or principles which are put at stake, and these are usually related to some underlying normative claims \citep{DBLP:conf/kdd/SrivastavaHK19,DBLP:journals/corr/abs-2010-10407}. Our research, however, indicates that these normative claims are often not unpacked and discussed to a sufficient extent in today's research on fairness in recommender systems. For instance, it may be argued that the issue with an e-commerce site optimizing for profit is not that it does so, but rather that it does so while misleading people into believing that recommendations are tailored to their needs. In situations such as this, the distinction between unfair and deceptive business activities can easily get blurred.}

\review{We note here that being fair to consumers or society in the bespoke examples may, in turn, also service providers,} e.g., when consumers establish long-term trust due to valuable recommendations or when they engage more with a music service when they discover more niche content. Finally, there are also~\emph{legal} guardrails that may come into play, e.g., when a large platform uses a monopoly-like market position to put certain providers inappropriately into bad positions. The current draft of the European Commission's Digital Service Act\footnote{\url{https://eur-lex.europa.eu/legal-content/en/TXT/?uri=COM:2020:825:FIN}} can be seen as a prime example where recommender systems and their potential harms are explicitly addressed in legal regulations\review{, as it~\dquotes{\textit{calls for more fairness, transparency and accountability for digital services’ content moderation processes, ensuring that fundamental rights are respected, and guaranteeing independent recourse to judicial redress}.}}

Overall, several examples exist where recommendations might be considered unfair for different stakeholders. In the context of the survey presented in this work, we are particularly interested in which specific~\emph{real-world} problems related to unfair recommendations are considered in the existing literature.

\subsection{Reasons for Unfairness}

There are different reasons why a recommender system might exhibit behavior that may be considered unfair. \review{For example, in~\cite{ekstrand2021fairness}, the authors report that unfairness can arise in many places, either in society, in the observations that form our data, and in the construction, evaluation, and application of decision support models. Similarly, in~\cite{DBLP:journals/ipm/AshokanH21}, the authors classify the biases in a computing system as pre-existing bias, technical bias, and emergent bias, whereas in~\cite{DBLP:journals/fdata/Olteanu00K19} the authors differentiate between issues introduced when collecting social data (in general, not focused on recommender systems), introduced while processing such data, pitfalls that occurred when analyzing data, and issues with the evaluation and interpretation of the findings. Herein, our discussions are based on insights from these and other earlier works, aiming to summarize and highlight the main causes of unfairness reported in the literature.}

One \review{first} common issue mentioned in the literature is that the data on which the machine learning model is trained is biased~\review{\citep{chen2020bias,DBLP:journals/fdata/Olteanu00K19}}. Such biases might, for example, result from the specifics of the data collection process, e.g., when a biased sampling strategy is applied.  A machine learning model may then ``pick up'' such a bias and reflect it in the resulting recommendations.

Another source of unfairness may lie in the machine learning model itself, e.g., when it reinforces existing biases or skewed distributions in the underlying data. Differences between recommendation algorithms in terms of reinforcing popularity biases and concentration effects were, for example, examined in~\cite{JannachLercheEtAl2015}. In some cases, the machine learning model might also directly consider a ``protected characteristic'' (or a proxy thereof) in its predictions~\citep{ekstrand2021fairness}. To avoid discrimination, and thus unfair treatment, of certain groups, a machine learning model should therefore not make use of protected characteristics such as age, color, or religion \review{(fairness through unawareness)~\citep{grgic2016case}}. \review{Despite its appealing simplicity, this definition has a clear issue, as sensitive characteristics may have historically affected non-sensitive characteristics (e.g., a person's GPA may have been influenced by their socioeconomic status). In order to adjust for biases in data collection or historical outcomes, it has been argued that, in fact, protected characteristics must be taken into account to place other observable features in context~\citep{KusnerCounterfactual2017}.}

Unfairness that is induced by the underlying data or algorithms may arise unknowingly to the recommendation provider. It is, however, also possible that a certain level of unfairness is designed into a recommendation algorithm, e.g., when a recommendation provider aims to maximize monetary business metrics while simultaneously keeping users satisfied as much as possible~\citep{ghanem2022balancing,JannachAdomaviciusVAMS2017}. Likewise, a recommendation provider may have a political agenda and particularly promote the distribution of information that mainly supports their own viewpoints.

Some works finally mention that the ``world itself may be unfair or unjust''~\citep{ekstrand2021fairness}, e.g., due to historical discrimination of certain groups. In the context of~\emph{algorithmic} fairness---which is the topic of our present work---such historical developments are, however, often not in the focus \review{even though the real reason certain characteristics are regarded protected is because of historical discrimination or subordination, where redress is necessary.} Rather, the question is to what extent this is reflected in the data or how this unfairness influences the fairness goals.

In general, the underlying reasons also determine \emph{where} in a machine learning pipeline\footnote{Consider~\cite{DBLP:journals/ipm/AshokanH21}, where the authors show that biases may occur in a typical machine learning pipeline from data generation, over the model building and evaluation, to deployment and user interaction.} interventions can or should be made to ensure fairness (or to mitigate unfairness). In a common categorization,~\citep{mehrabi2021fairnesssurvey,Shrestha19Fairness,pitoura2021fairness,zehlike2021fairness}, this could be achieved
\emph{(i)} in a data pre-processing phase, \emph{(ii)} during model learning and optimization, and \emph{(iii)} in a post-processing phase. In particular, in the model learning and post-processing phase, fairness-ensuring algorithmic interventions must be guided by an \emph{operationalizable} (i.e., mathematically expressed) goal. In the case of affirmative action policies,
one could, for example, aim to have an equal distribution of recommendations of members of the majority group and members of an underrepresented group.
As we will see in Section~\ref{sec:landscape},
such a goal is often formalized as a target distribution and/or as an evaluation metric to gauge the level of existing or mitigated fairness.

\subsection{Notions of Fairness}
\label{ss:notions}

When dealing with phenomena of unfairness such as those outlined, and when our purpose is to prevent or mitigate such phenomena, a question arises: what do we consider fair in general and in a particular application context? 
Fairness, in general, is fundamentally a societal construct or a~\emph{human value}, which has been discussed for centuries in many disciplines like philosophy and moral ethics, sociology, law, or economics. Correspondingly, countless definitions of fairness were proposed in different contexts, see for example~\citep{Verma18definitions,Verma20facets} for a high-level discussion of the definition of fairness in machine learning and ranking algorithms, or~\cite{Mulligan2019ThisThing} for the relationship to social science conception of fairness. \review{As we will see in the remainder of this survey, fairness is a complex concept with multiple perspectives. Consequently, there are numerous definitions, but none of them appear to be exhaustive.} 

\review{In general, the societal constructs around fairness mainly depend on how moral standards or dilemmas are addressed: either through~\textit{descriptive} or~\textit{normative} approaches~\citep{DBLP:conf/kdd/SrivastavaHK19}. While normative ethics involves creating or evaluating moral standards to decide what people should do or whether their current moral behavior is reasonable, descriptive (or comparative) ethics is a form of empirical research into the attitudes of individuals or groups of people towards morality and moral decision-making. As mentioned above, normative claims are often not explicitly specified in existing research, both in general machine learning and in recommender systems research. In fact, it was already recommended in earlier research to make these assumptions more explicit~\citep{DBLP:journals/corr/abs-2010-10407}. From our study of the literature, we observe that a majority of the works did not clarify what the actual normative claim is being addressed or who is representing or making such claims.} 

\review{As a possible consequence of this problem, we also observe that researchers, in most cases, do not refer to a specific public discussion of the topic at hand. For many papers on recommender systems, there is, for example, no indication or evidence that there is a public debate outside computer science, e.g., whether or not it is fair to recommend niche movies. Nonetheless, it is true that there actually are areas, like job recommendation, where a public discussion takes place, e.g., about discrimination and what normative claims are agreed to be addressed.}

\review{The primary notions of fairness that will be used throughout this review---as extracted from the aforementioned literature and recent surveys~\citep{Li2022Survey,Wang2022ASurveyTOIS}---are presented next and further expanded in Section~\ref{ss:landscape}. We emphasize that these definitions present a specific perspective on defining the concept of fairness. They are, however, not necessarily~\textit{orthogonal} and~\textit{all-encompassing}. Table~\ref{tab:fairness-notions-examples} shows examples of fictitious statements of a user regarding unfairness in a job recommendation scenario under different notions of fairness.}

\begin{itemize}
    \item \review{\emph{Group vs.~individual}: Individual fairness roughly expresses that similar individuals should be treated similarly, e.g., candidates with similar qualifications should be ranked similarly in a job recommendation scenario. Group fairness, in contrast, aims to ensure that ``\review{different groups} have similar experience''~\citep{ekstrand2021fairness}, i.e., protected groups receive similar benefits from the decision-making as others. Typical groups in such a context are a majority or dominant group and a protected group (e.g., an ethnic minority). Since this may be too simplistic, other authors state \textit{we are all equal} as the fundamental logic underlying group fairness \citep{DBLP:journals/cacm/FriedlerSV21}, asserting their equivalence as a starting point.}
    \item \review{\emph{Process vs.~outcome:} Process (or: treatment) unfairness means that individuals with similar non-sensitive attributes receive different outcomes solely due to the difference in sensitive features. Outcome (or: impact) unfairness occurs when a system produces outputs that benefit (harm) a group of individuals sharing a sensitive attribute value more frequently than other groups~\citep{zafar2017fairness}. Put it differently, process fairness assesses aspects such as the data used, the decision-making principles of the system, and the causal association between inputs and outputs. In contrast, outcome fairness disregards the internal operation of the system and concentrates solely on the equitable distribution of rewards~\citep{amigo2022unifying}.}
    \item \review{\emph{Direct vs.~indirect}: Fairness can also be analyzed based on whether particular sensitive feature holders are directly harmed or not \citep{national2004measuring}. Direct fairness refers to situations in which persons receive less favorable treatment based on protected characteristics such as race, religion, or gender. When the reasons for the discrimination are only tenuously connected to (or identical to) the protected characteristic, we have indirect fairness.\footnote{\review{The term \textit{redlining}~\citep{CorbettDavies2018Mismeasure} is analogous to the concept of indirect unfairness wherein a non-sensitive characteristic (such as geography) is used as a proxy for a more personal quality (such as race or socioeconomic status).}} For example, some institutions use the location of candidates as a~\textit{proxy} for an overtly discriminating characteristic (e.g., race)~\citep{zhang2018fairness}.}
    \item \review{\emph{Statistical vs.~predictive parity}: In machine learning, fairness definitions fundamentally seek some sort of equity on various portions of the~\textit{confusion matrix} used for binary classification evaluation. Statistical parity is independent of the actual value and requires protected group members to have an equal positive prediction rate. Predictive parity employs the actual outcome and requires that the model's precision (or accuracy) is comparable for all subgroups under consideration.} 
    \item \review{\emph{Static vs.~dynamic}: In static fairness, the recommendation environment is fixed during the recommendation process; hence, the user activity level is assumed to remain unchanged. Dynamic fairness definitions, on the other hand, integrate the (typical) dynamic attribute of most recommender systems, which needs to consider new user interactions, new items, or continually evolving user groups.}
    \item \review{\emph{Associative vs.~causal}: Associative fairness metrics are computed based on data and do not allow reason about the causal relations between the features and the decisions. Causal fairness definitions, on the other hand, are usually defined in terms of (non-observable) interventions and counterfactuals and tend to consider the additional structural knowledge of the system regarding how variables propagate on a causal model \citep{Li2022Survey}.}
\end{itemize}

\begin{table}[h!tb]
\caption{\review{Examples for possible statements around different notions of fairness in the context of a recommender system for jobs.}}
\label{tab:fairness-notions-examples}
\centering
\color{black}
\begin{tabular}{p{5cm}cp{5cm}}
\toprule
  \textbf{Group} & vs.~& \textbf{Individual}\leavevmode\leavevmode\\
  \emph{Compared to men, women are recommended low-paying occupations! } & & \emph{My friend Elisa and I had similar GPA, qualifications and skills, but she got better job recommendations!} \leavevmode\\  \rule{-2pt}{5ex}
  \noindent \textbf{Process} & vs.~& \textbf{Outcome}\leavevmode\\
  \emph{My friend John and I had similar GPA, qualifications, and skills, but he got better suggestions only because he's a man!} & & \emph{Relevant higher-paying jobs get recommended to white people rather than black!} \leavevmode\\  \rule{-2pt}{5ex}
  \textbf{Direct} & vs.~& \textbf{Indirect}\leavevmode\\
  \emph{I am receiving worse recommendations only because of my skin color!} & & \emph{People from south Italy receive worse job recommendations by the system!}\leavevmode\\
  \rule{-2pt}{5ex}
  \textbf{Statistical parity} & vs.~& \textbf{Predictive parity} \\
  \emph{My group should receive as many good recommendations as other groups!} & & \emph{Among people who are recommended for the job, there is a smaller share of qualified people from my group than from other groups!} \\ \rule{-2pt}{5ex}
  \textbf{Static} & vs.~& \textbf{Dynamic} \\
  \emph{The system achieved to be fair just once, in a different job market, but now employees' goals and priorities have changed!} & & \emph{The system accounts for shifts in our tastes and needs, and can prefer me today if it preferred someone else yesterday!}\leavevmode\\ \rule{-2pt}{5ex}
  \textbf{Associative} & vs.~& \textbf{Causal}\leavevmode\\
  \emph{If you are black-skinned, you are historically more likely to be discriminated against!} & & \emph{Had I not been black-skinned, would I have received that recommendation?}\leavevmode\\
  \bottomrule
\end{tabular}
  \color{black}

\end{table}

\review{\noindent Other categorizations can be found in the literature, based on~\emph{short-term vs.~long-term} considerations (according to the duration of the fairness requirements),~\emph{granularity} (whether the system applies the same fairness notion to everyone or if users could decide how they want to be treated by the system),~\emph{transparency} (to discriminate notions that are explainable from those that are a black box), or~\emph{depending on the associated fairness concept} (such as consistent, calibrated, counterfactual, Rawlsian maximin, envy-free, and maximin-shared)~\citep{Li2022Survey,Wang2022ASurveyTOIS,amigo2022unifying}.}
An in-depth discussion of these--sometimes even incompatible \review{\citep{Verma18definitions,amigo2022unifying}}--notions
of fairness is beyond the scope of this work, which focuses on an analysis of how scholars in recommender systems operationalize the research problem. For questions of individual fairness, this might relate to the problem of defining a similarity function. For certain group fairness goals, on the other hand, one has to determine which are the (protected) attributes that determine group membership. Furthermore, it is often required to define/indicate
precisely some \emph{target distributions}. Later, in Section \ref{sec:landscape}, where we review the current literature, we will introduce additional notions of fairness and their operationalizations as they are found in the studied papers. As we will see, a key point here is that researchers often propose to use very abstract operationalizations (e.g., in the form of fairness metrics), which was identified earlier as a potential key problem in the broader area of fair ML in \cite{selbst2019}.

\subsection{Related Concepts: Responsible Recommendation and Biases}
Issues of fairness are often discussed within the broader area of \emph{responsible} recommendation~\citep{elahi2021aiethics,ekstrand2021fairness,di2022recommender}, with the key dimensions \textit{generalizability},~\textit{robustness} \citep{deldjoo2020adversarial,deldjoo2021survey},~\textit{privacy}~\citep{anelli2021pursuing,friedman2015privacy},~\textit{interpretability}~\citep{tintarev2022beyond,deldjoo2022review}, and \textit{fairness}, with the definitions of these concepts blurring as we progress through the list. In \cite{elahi2021aiethics}, the authors, in particular, discuss the potential negative effects of recommendations and their underlying reasons with a focus on the media domain. Specific phenomena in this domain include the emergence of filter bubbles and echo chambers. There are, however, also other more general potential harms such as popularity biases as well as fairness-related aspects like
discrimination 
that can emerge in media recommendation setting\review{, for example, when one gender or race is treated differently just based on this attribute, as when suggesting images for a specific profession}.
Fairness is therefore seen as a particular aspect of responsible recommendation in~\cite{elahi2021aiethics}. A similar view is taken in~\cite{ekstrand2021fairness}, where the authors review a number of related concerns of responsibility: accountability, transparency, safety, privacy, and ethics. In the context of our present work, most of these concepts are however only of secondary interest.

More important, however, is the use of the term \emph{bias} in the related literature.
As discussed above, one frequently discussed topic in the area of recommender systems is the problem of \emph{biased data} \citep{chen2020bias,Baeza2018CACM}. One issue in this context is that the data that is collected from existing websites---e.g., regarding which content visitors view or what consumers purchase---may \review{in part be the result of an already existing recommender system and, hence,}
biased by what is shown to users.
This, in turn, then may lead to biased recommendations when machine learning models reflect or reinforce the bias, as mentioned above. In works that address this problem, the term bias is often used in a more statistical sense, as done in~\cite{ekstrand2021fairness}. However, the use of the term is inconsistent in the literature, as also observed in in~\cite{chen2020bias} and in our work. In some early papers, bias is used almost synonymously with fairness.  In \cite{Friedman1996Bias}, for example, bias is used to ``\emph{refer to computer systems that systematically and unfairly discriminate against certain individuals or groups of individuals in favor of others}''. In our work, we acknowledge that biased recommendations may be unfair, but we do not generally equate bias with unfairness. Considering the problem of popularity bias in recommender systems, such a bias may lead to an over-proportional exposure of certain items to users. This, however, not necessarily leads to unfairness in an ethical or legal sense\review{. Instead, it all depends on the underlying ethical principles and normative claims, as discussed before}.
\review{Moreover, an in-depth discussion and systematic comparison of various forms of biases is beyond the scope of our work; we instead refer the reader to~\cite{chen2020bias}, where different forms of biases are discussed in more depth.}

\section{Research Methodology}
\label{sec:methodology}
In this section, we first describe our methodology for identifying relevant papers for our survey. Afterward, briefly discuss how our survey extends previous works in this area.

\subsection{Paper Collection Process}
\review{We adopted a mixed and semi-systematic approach to identify relevant research papers.\footnote{\review{We note here that our work is not intended to be a systematic literature review in the strict sense of \cite{DBLP:journals/infsof/KitchenhamBBTBL09}, but rather aims to outline a broader picture of current research activities.}}} \review{In the first step, we identified relevant research papers by querying the DBLP\footnote{\review{\url{https://dblp.org/}}} digital library with predefined search terms and a set of explicit criteria for inclusion and exclusion.}
\review{Afterwards, to include relevant papers which did not match the search terms in this still-evolving field, we (a) applied a snow-balling procedure and (b) relied on researcher experience to identify other relevant papers that were published in focused outlets.}

Based on our prior knowledge about the literature, we used the following search terms in order to cover a wide range of works in an emerging area, where terminology is not yet entirely unified: \emph{fair recommend}, \emph{fair collaborative system}, \emph{fair collaborative filtering}, \emph{bias recommend}, \emph{debias recommend}, \emph{fair ranking}, \emph{bias ranking}, \emph{unbias ranking}, \emph{re-ranking recommend}, \emph{reranking recommend}.
To identify papers, we queried DBLP in its respective search syntax, stating that the provided keywords must appear in the title of the paper. 

From the returned results, we then removed all papers that were published \review{only} as preprints on arXiv.org\footnote{Note that DBLP indexes arXiv papers.} and we removed survey papers. We then manually scanned the remaining \review{268} papers. In order to be included in this survey, a paper had to fulfill the following additional criteria:
\begin{itemize}
  \item It had to be explicitly about \emph{fairness}, at least by mentioning this concept somewhere in the paper. Papers which, for example, focus on mitigating popularity biases, but which do not mention that fairness is an underlying goal of their work, were thus not considered.
  \item It had to be about \emph{recommender systems}. Given the inclusiveness of our set of query terms, a number of papers were returned which focused on fair information retrieval. Such works were also excluded from our study.
\end{itemize}

This process left us with \review{157} papers. The papers were read by at least two researchers and categorized in various dimensions, see Section~\ref{sec:landscape}.\footnote{The full list of papers is made publicly available in this link:~\url{https://github.com/yasdel/FairnessRecSys_Survey2023}.}

\subsection{Relation to Previous Surveys}\label{ss:prev}
A number of related surveys were published in the last few years.  The survey provided by~\cite{chen2020bias} focuses on biases in recommender systems, and connects different types of biases, e.g., popularity biases, with questions of fairness, see also \citep{Abdollahpouri2020Connection}.
Note that bias mitigation in recommendation mostly focuses on increasing the accuracy or robustness of the recommendations through debiasing approaches, rather than on promoting fairness.

The recent monograph by~\cite{ekstrand2021fairness} discusses fairness aspects in the broader context of \emph{information access} systems, an area that covers both information retrieval and recommender systems. Their comprehensive work in particular includes a taxonomy of various fairness dimensions, which also serves as a foundation of our present work. 
This study differs from our work in that our objective is not to give a fresh classification of fairness concepts and methods found in the literature.  Instead, our main objective is to investigate the current state of existing research, e.g., in terms of which concepts and algorithmic approaches are predominantly investigated and where there might be research gaps.
\review{Ekstrand et al., on the other hand, focus more generally on future directions in this area.}

Different survey papers were published also in the more general area of fair machine learning or fair AI, as mentioned above \citep{mehrabi2021fairnesssurvey,barocas-hardt-narayanan}. Clearly, many questions and principles of fair AI apply also to recommender systems, which can be seen as a highly successful area of applied machine learning. Differently from such more general works, however, our present work focuses on the particularities of fairness in recommender systems.

Very recently, while we conducted our research, a number of alternative surveys on fairness in recommender systems have become available as preprints or peer-reviewed publications, including \cite{pitoura2021fairness}, \cite{Zehlike2022part2}, \cite{Wang2022ASurveyTOIS}, and \cite{Li2022Survey}. Clearly, there is a certain overlap of our survey and these recent publications, e.g., in terms of the used taxonomy of fairness-related aspects. Note, however, that unlike some of these papers, e.g., \cite{Li2022Survey,pitoura2021fairness}, our aim is \emph{not} to establish a new taxonomy or to discuss the technical details of specific computational metrics or algorithmic approaches that were proposed in the past literature. Instead, our aim is to paint a landscape of existing research and to thereby identify potential research gaps. In that context, our work has similarities with the work by~\cite{Wang2022ASurveyTOIS}, who reviewed and categorized 60 recent works on fairness in recommender systems. While our survey involves a larger number of papers, Wang et al.~dive deeper into the technicalities of particular approaches, which is not the focus of our work.
Here, in contrast, we aim to paint a broader picture of today's research activities and existing gaps without entering into the technical specifics of existing approaches. Moreover, our work also emphasizes more on evaluation aspects  and on potential methodological issues in this research area.  The recent work by~\cite{Zehlike2022part2}, finally, mainly discusses individual research works in detail, also including more general ones on learning-to-rank. The overlap with this work, except for the discussion of different dimensions of fairness, is therefore limited.

\review{In general, the goal of these existing works is mainly to review and synthesize the various existing approaches so far to design fair recommender systems and to evaluate them. The goal of our work is indeed different, as we aim to analyze and quantify which notions of fairness the research community is working on and how the research problem is operationalized. Differently from previous surveys, our study can therefore inform about the less frequently studied areas, and thus potential gaps, of fairness research in a quantitative manner. Moreover, our analyses of the applied research methodologies reveal a very strong predominance of data-based experiments, which rely on abstract computational metrics and do not involve humans in the loop. We, therefore, believe that our survey complements existing surveys well.}

\color{black}

\section{Landscape of \review{Fairness} Research \review{in Recommender Systems}}
\label{sec:landscape}
In this section, we categorize the identified literature along different dimensions to paint a landscape of current research and to identify existing research gaps.

\subsection{Publication Activity per Year}
\begin{figure*}[htp]
     \centering
     \includegraphics[width=0.6\linewidth]{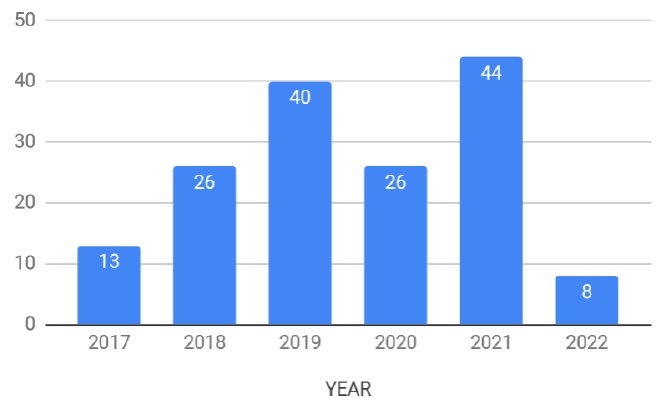}
     \caption{Number of papers published per year. \review{The entire number of papers sum up to 157.} }
     \label{fig:historical-development}
 \end{figure*}

Interest in fairness in recommender systems has been constantly growing over the past few years.
Figure~\ref{fig:historical-development}
shows the number of papers per year that were considered in our survey. Questions of fairness in information retrieval have been discussed for many years,
see, e.g., \cite{Pedreshi2008Distriminationaware} for an earlier work. \review{The area has been consistently growing since then, leading also to the establishment of dedicated conference series like the ACM Conference on Fairness, Accountability, and Transparency (ACM FAccT).\footnote{\review{A number of related events have been recently connected through the ACM FAccT Network, \url{https://facctconference.org/network/}}}}
In the area of recommender systems, however, the earliest paper we identified through our search, which only considers papers in which fairness is \emph{explicitly} addressed, was published as late as in 2017.

\subsection{Types of Contributions} Academic research on recommender systems in general is largely dominated by algorithmic contributions, and we correspondingly observe a large amount of new methods that are published every year. Clearly, building an effective recommender system requires more than a smart algorithm, e.g., because recommendation to a large extent is also a problem of human-computer interaction and user experience design \citep{JannachResnickEtAl2016,jannach2021aimagintro}. Now when questions of fairness should be considered as well, the problem becomes even more complex as for example ethical questions may come into play and we may be interested on the impact of recommendations on individual stakeholders, including society.

In the context of our study, we were therefore interested in which \emph{general types} of contributions we find in the computer science and information systems literature on fair recommendation. Based on the analysis of the relevant papers, we first identified two general types of works: (a) \emph{technical} papers, which, e.g., propose new algorithms, protocols, and metrics or analyze data, and (b) \emph{conceptual} papers. The latter class of papers is diverse and includes, for example, papers that discuss different dimensions of fair recommendations,
papers that propose conceptual frameworks, or works that connect fairness with other quality dimensions like diversity.

We then further categorized the technical papers in terms of their \emph{specific technical type} of contribution. The main categories we identified \review{based on the research contributions of the surveyed papers} are (a) \emph{algorithm} papers, which for example propose re-ranking techniques, (b) \emph{analytic} papers, which for example study the outcomes of a given algorithm, and (c) \emph{methodology} papers, which propose new metrics or evaluation protocols.

\begin{figure*}[h!t]
     \centering
     \includegraphics[width=0.45\linewidth]{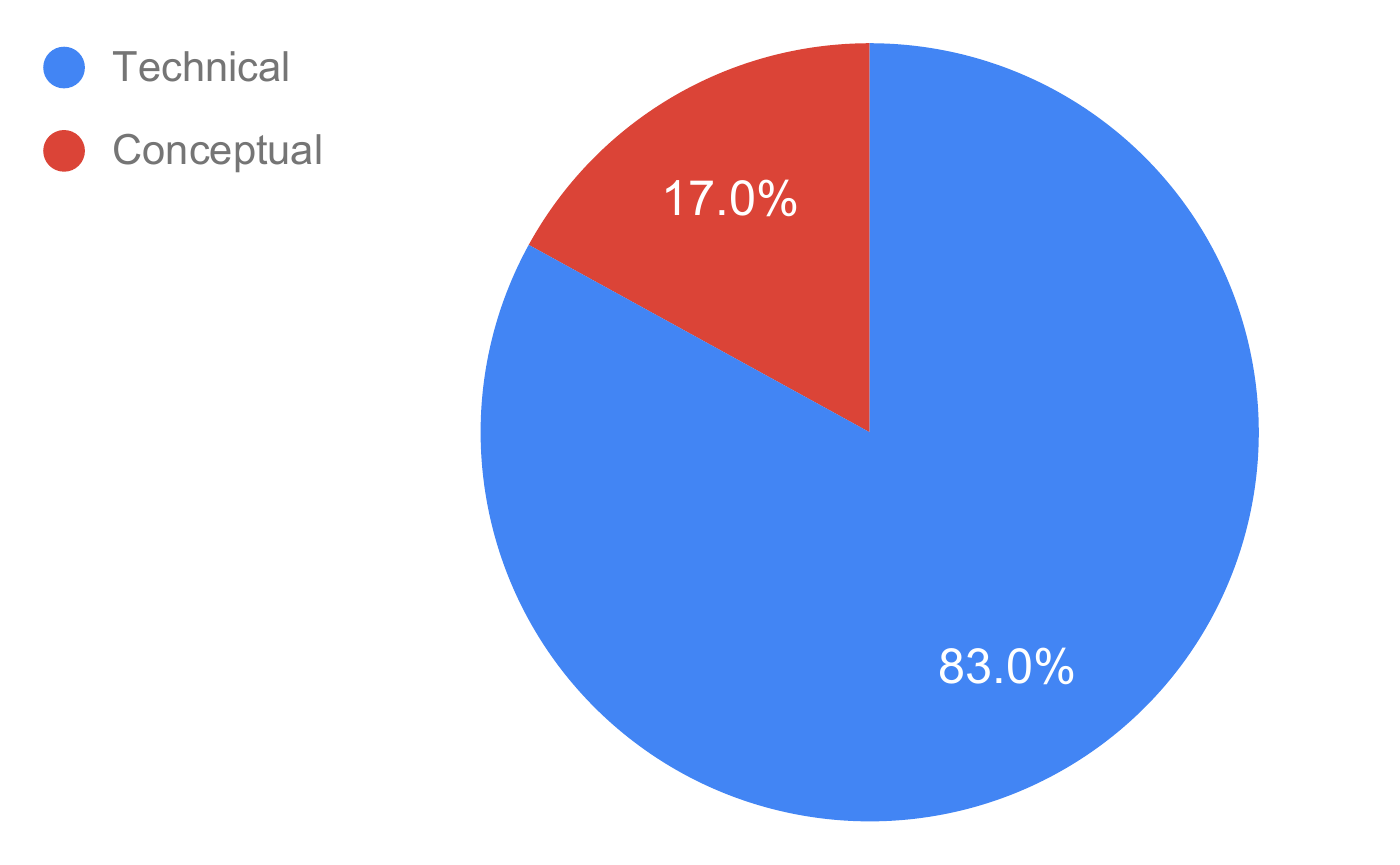}
     \caption{Technical vs.~Conceptual Papers.}
     \label{fig:technical-vs-conceptual}
 \end{figure*}

Figure~\ref{fig:technical-vs-conceptual} shows how many papers in our survey were considered as technical and conceptual papers.
Non-technical papers cover a wide range of contributions, such as guidelines for designers to avoid \textit{compounding} previous injustices \citep{DBLP:conf/um/Schelenz21}, exploratory studies that investigate user perceptions of fairness \citep{Sonboli2021Fairness}, or discussions about how difficult it is to audit these types of systems \citep{DBLP:conf/bias/KrafftHZ20}.

We observe that today's research on fairness on recommender systems is dominated by technical papers. In addition, we find that the majority of these works focuses on improved algorithms, e.g., to debias data or to obtain a fairer recommendation outcome through list re-ranking. To some extent this is expected as we focus on the computer science literature. However, we have to keep in mind that the concepts of fairness and unfairness or social constructs may depend on a variety of environmental factors in which a recommender system is deployed. As such, the research focus in the area of fair recommender systems seems rather narrow and on algorithmic solutions. As we will observe later, however,
such algorithmic solutions commonly assume that some pre-existing and mathematically defined optimization goals are available, e.g., a target distribution of recommendations. In practical applications, the major challenges
mostly lie (a) in establishing a common understanding and agreement on such \review{fairness goals} and (b) in finding or designing operationalizable optimization goals (e.g., a computational metric) which represent reliable measures or proxies for the given fairness goals.

\subsection{Categorization of Notions of Fairness in Literature}
\label{subsec:notions}

In \cite{Li2021tutorial}, a taxonomy of different notions of fairness was introduced\review{: group vs.~individual, single-sided vs.~multi-sided, static vs.~dynamic, and associative vs.~causal fairness; see also our discussions in Section~\ref{ss:notions}.} In the following, we review the literature following this taxonomy.\footnote{\review{Each paper was categorized by at least two researchers, and potential discrepancies were resolved through a discussion process. The same process was applied to categorize the papers also in other dimensions as discussed later in this section.}}

\paragraph{Group vs.~Individual Fairness}
A very common differentiation in fair recommendation is to distinguish between \emph{group} fairness and \emph{individual} fairness, as indicated before.
With group fairness, the goal is to achieve some sort of \emph{statistical parity} between \emph{protected} groups \citep{Binns2019Apparent}.  In fair machine learning, a traditional goal often is to ensure that there are equal number of members of each protected group in the outcome, e.g., when it comes to make a ranked list of job candidates. The protected groups in such situations are commonly determined by characteristics like age, gender, or ethnicity. Achieving individual fairness in the described scenario means that candidates with similar characteristics should be treated similarly. To operationalize this idea, therefore some distance metric is needed to assess the similarity of individuals. This can be a challenging task, since there is no consensus on the notion of similarity, and it could be task-specific \citep{Dwork12}.
Ideas of individual fairness in machine learning were discussed in an early work in \cite{Dwork12}, where it was also observed that achieving group fairness might lead to an unfair treatment at the individual level. In the candidate ranking example, favoring members of protected groups to achieve parity might ultimately result in the non-consideration of a better qualified candidate from a non-protected group. As a result, group and individual fairness are frequently viewed as trade-offs, which is not always immediately evident \citep{Binns2019Apparent}.

\begin{figure*}[tb]
     \centering
     \includegraphics[width=0.6\linewidth]{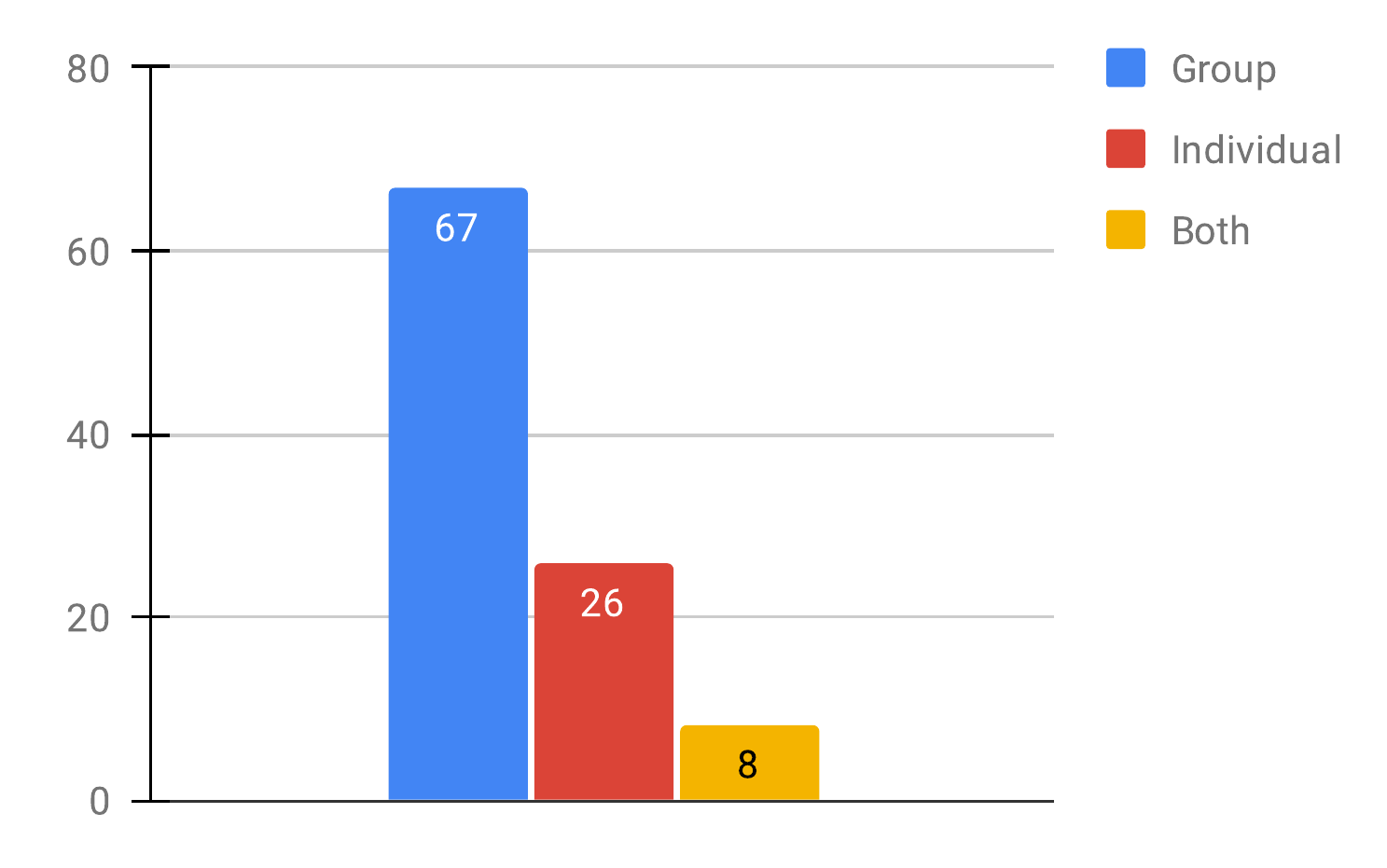}
     \caption{Group vs.~Individual Fairness.}
     \label{fig:individual-vs-group-fairness}
\end{figure*}

Figure~\ref{fig:individual-vs-group-fairness} shows how many of the surveyed papers focus on each category.
The figure shows that research on scenarios where group fairness is more common than works that adopt the concept of individual fairness. Only in rare cases, both types of fairness are considered.

Group fairness entails comparing, on average, the members of the privileged group against the unprivileged group. One overarching aspect to identify research papers on groups fairness is the distinction between the \textit{(i)} benefit type (exposure vs.~relevance), and \textit{(ii)} major stakeholders (consumer vs.~provider).  Exposure relates to the degree to which items or item groups are exposed uniformly to all users/user groups. Relevance (accuracy) indicates how well an item's exposure is effective, i.e., how well it meets the user's preference. For recommender systems, where users are first-class citizens, there are multiple stakeholders, consumers, producers, and \review{other stakeholders} (see next section).

To perform fairness evaluation for item recommendation tasks, the users or items are divided into non-overlapping groups (segments) based on some form of \textit{attributes}. These attributes can be either supplied externally by the data provider (e.g., gender, age, race) or
computed internally\review{\footnote{\review{We should note that we found no example where the reliability of these implicitly computed attributes was analyzed. Usually, authors use explicit thresholds to assign users/items to groups \citep{Li2012UserOriented,Xiao20SocialActiv} or percentiles from distributions based on a variable of interest, such as item popularity \citep{Abdollahpouri21User,deldjoo2021flexible}.}}} from the interaction data (e.g., based on user activity level, mainstreamness, or item popularity)\review{ \citep{Abdollahpouri21User,Li2012UserOriented}}.
In Table~\ref{tab:goals}, we provide a list of the most commonly used attributes in the recommendation fairness literature, which can be utilized to operationalize the group fairness concept. They are divided according to Consumer fairness (C-Fairness), Producer Fairness (P-Fairness), and combinations (CP-Fairness)~\review{\citep{Burke2017Fairness} or multi-sided fairness.}.

\review{Additionally, it is possible to observe in RS settings that these sensitive attributes may be provided by external providers as demographic metadata (for example, user's gender, age, occupation), or they may be extracted from user-item interaction data, for example, dividing users based on their level of activity (i.e., active vs. inactive users), or the types of items they consume (e.g., mainstream-users vs. non-mainstream). Here a related concept is \textit{obfuscation}~\citep{slokom2021towards}, which is a strategy for privacy protection to conceal sensitive information. Fairness and privacy can be considered as interwoven under obfuscation, as described by~\cite{Dwork12} and \cite{pessach2022review}, where a violation of privacy can lead to unfairness due to an adversary's capacity to infer sensitive information about an individual and utilize it in a discriminatory manner.}

\afterpage{\clearpage
\begin{table*}[!h]
\caption{Overview of common attributes used when addressing fairness concepts from consumers, providers, or both perspectives.}
\label{tab:goals}
\footnotesize
\begin{center}
\begin{tabularx}{1.10\textwidth}[t]{XX}
\hline
\textbf{\textcolor{black}{Goal 1: Consumer Fairness}} & Attribute\\
\hline
Target: \textit{Demographic parity} -- sensitive attributes are attained by birth and not under a user's control. &
\begin{minipage}[t]{\linewidth}%
\begin{itemize}
\item Gender \citep{deldjoo2021flexible,deldjoo2021explaining,Wu21decomposed,DBLP:conf/icml/GorantlaDL21,wang2021practical,ghosh2021fair,wan2020addressing,edizel2020fairecsys,Tsintzou19BiasDisparity,DBLP:conf/recsys/MansouryMBP19,DBLP:conf/recsys/LinSMB19,DBLP:conf/kdd/GeyikAK19,DBLP:journals/kbs/XiaYXL19,DBLP:conf/fat/BurkeSO18,DBLP:conf/icwsm/ChakrabortyMBGG17,Farnadi2018AFairness,burke2017balanced,riederer2017price}
\item Race \citep{DBLP:conf/icml/GorantlaDL21,ghosh2021fair,DBLP:conf/um/ZhengDMK18,DBLP:conf/cikm/ZhuHC18,DBLP:conf/icwsm/ChakrabortyMBGG17,DBLP:journals/corr/abs-1809-03040,riederer2017price}
\item Age \citep{deldjoo2021flexible,DBLP:journals/ijimai/BobadillaLG021,suhr2021does,DBLP:journals/ipm/MelchiorreRPBLS21,DBLP:conf/icml/GorantlaDL21,Farnadi2018AFairness}
\item Nationality \citep{Weydemann19location} and Location \citep{riederer2017price}
\item Occupation \citep{Farnadi2018AFairness}
\vspace{2mm}
\end{itemize}
\end{minipage}\leavevmode\\ [1ex]

Target: \textit{Merit-based fairness} -- attained through a user's merit over time.&
\begin{minipage}[t]{\linewidth}%
\begin{itemize}
\item Education \citep{suhr2021does,GomezWinner2021}
\item Income \citep{suhr2021does}
\end{itemize}
\end{minipage}\leavevmode\\ [1ex]

Target: \textit{Behavior-oriented fairness}  --  attained based on a user's engagement with the system/item catalog.&
\begin{minipage}[t]{\linewidth}%
\begin{itemize}
\item User (in)activeness \citep{Hao21Pareto,Li2012UserOriented,Xiao20SocialActiv,Fu20Explainable,Chakraborty19Equality}
\item User (non)mainstreamness \citep{Abdollahpouri2020Connection,Abdollahpouri21User}
\end{itemize}
\end{minipage}\leavevmode\\ [1ex]

Target: \textit{Other emerging attributes}
&
\begin{minipage}[t]{\linewidth}%
\begin{itemize}
\item Physio/psychological \citep{wan2020addressing,htun2021perception}
\item Sentiment-based \citep{Lin21sentiment}
\end{itemize}
\end{minipage}\leavevmode\\ [2pt]
\hline
\textbf{\textcolor{black}{Goal 2: Provider Fairness}}\leavevmode\\
\hline

Target: \textit{Item producer/creator}  -- sensitive attribute based on who the item producer is.&
\begin{minipage}[t]{\linewidth}%
\begin{itemize}
\item News author \citep{GharahighehiVP21}, music artist \citep{Ferraro19Music}, movie director \citep{Boratto21Interplay}
\end{itemize}
\end{minipage}\leavevmode\\
Target \textit{Producer's demographic or general information}  -- sensitive attribute based on to which demographic group the item producer belongs, e.g., male vs.~female artists. &
\begin{minipage}[t]{\linewidth}%
\begin{itemize}
\item Gender \citep{Kirnap21Estimation,Boratto21Interplay,Shakespeare20Exploring,DBLP:journals/kbs/XiaYXL19}, geographical region \citep{GomezWinner2021}
\end{itemize}
\end{minipage}\leavevmode\\
Target: \textit{Item information}  -- sensitive attribute based on the item information itself. &
\begin{minipage}[t]{\linewidth}%
\begin{itemize}
\item Price and brand  \citep{deldjoo2021flexible,Dash21umpire}, geographical region \citep{DBLP:conf/pakdd/LiuLTLCH20,DBLP:conf/fat/BurkeSO18}
\end{itemize}
\end{minipage}\leavevmode\\
Target: \textit{Interaction-oriented fairness}  -- sensitive attribute based on the interactions observed on items e.g., popularity. &
\begin{minipage}[t]{\linewidth}%
\begin{itemize}
\item Popularity \citep{deldjoo2021flexible,DBLP:journals/access/DongXL21,DASILVA2021115112,Wundervald21Cluster,Borges21mitigating,Ge21Towards,Sun19Debiasing,Weydemann19location,Abdollahpouri19TheUnfairness,Zhu18FMSR}, cold items \citep{Zhu21NewItems}
\end{itemize}
\end{minipage}\leavevmode\\
Target: \textit{Other emerging attributes}  &
\begin{minipage}[t]{\linewidth}%
\begin{itemize}
\item Premium membership \citep{Deldjoo19GCE}, sentiment and reputation \citep{Lin21sentiment,Zhu20FARM}
\end{itemize}
\end{minipage}\leavevmode\\
Target: \textit{Non-sensitive attributes}  &
\begin{minipage}[t]{\linewidth}%
\begin{itemize}
\item Movie and music genre \citep{Tsintzou19BiasDisparity,DBLP:conf/recsys/LinSMB19,rastegarpanah2019fighting,Ferraro19Music}
\end{itemize}
\end{minipage}\leavevmode\\ \hline
\multicolumn{2}{l}{\textbf{\textcolor{black}{Goal 3: Consumer Provider Fairness (Multi-sided Fairness)}}}\leavevmode\\
\hline
Target: Combinations of two targets from C-Fairness and P-Fairness.&
\begin{minipage}[t]{\linewidth}%
\begin{itemize}
\item Same category of sensitive attributes for both users and items (e.g. \textit{behavior-oriented}) \citep{naghiaei2022cpfair,rahmani2022role,rahmani2022unfairness,Lin21sentiment,Abdollahpouri19TheUnfairness,DBLP:conf/fat/BurkeSO18}
\item Different categories of sensitive attributes \citep{deldjoo2021flexible,Deldjoo19GCE,DBLP:conf/recsys/MansouryMBP19,Tsintzou19BiasDisparity,Weydemann19location,DBLP:journals/kbs/XiaYXL19,rahmani2022experiments}
\end{itemize}
\end{minipage}\leavevmode\\ [0.5ex]
\hline
\end{tabularx}
\end{center}
\end{table*}
\clearpage}

Moreover, in the area of recommender systems, a number of \emph{people recommendation} scenarios can be identified that are similar to classical fair ML problems. These include recommenders on dating sites, social media sites that provide suggestions for connections, and specific applications, e.g., in the educational context \citep{GomezWinner2021}. In these cases, user demographics may play a major role\review{, together with other factors such as popularity, expertise, and availability at a certain point in time}.
However, in many other cases, e.g., in e-commerce recommendation or media recommendation, it is not always immediately clear what protected groups may be. In
\cite{Li2012UserOriented} and other works, for example, user groups are defined based on their activity level, and it is observed that highly active users (of an e-commerce site) receive higher-quality recommendations in terms of usual accuracy measures. This is in general not surprising because there is more information a recommender system can use to make suggestions for more active users. However, it stands to question if an algorithm that returns the best recommendations it can generate given the available amount of information should be considered unfair \review{per se}.
\review{In fact, merely observing different levels of recommendation accuracy for more active and less active users may not be enough to conclude that a system is unfair. Instead, it is important to carefully elaborate on the underlying reasons and the related normative claims. Some particular user groups may for example have had fewer opportunities to engage with a system.}

Recent studies have also focused on two-sided CP-Fairness, as illustrated in \cite{naghiaei2022cpfair,rahmani2022unfairness}. In these works, the authors demonstrate the existence of inequity in terms of exposure to popular products and the quality of recommendation offered to active users. It is unknown if increasing fairness on one or both sides (consumer/producers) has an effect on the overall quality of the system. In \cite{naghiaei2022cpfair}, an optimization-based re-ranking strategy is then presented that leverages consumer and provider-side benefits as constraints. The authors demonstrate that it is feasible to boost fairness on both the user and item sides without compromising (and even enhancing) recommendation quality.

Different from traditional fairness problems in ML, research in fairness for recommenders also frequently considers the concept of \emph{fairness towards items} or their providers (suppliers), see also \citep{Li2021tutorial}, which differentiates between user and item fairness. \review{In these research works, the idea often is to avoid an unequal (or: unfair) \emph{exposure} of items
from different providers, e.g., artists in a music recommendation scenario. The term \emph{item fairness}, although used in the literature, may however not be optimal. In reality, it might be argued that this perspective is only important because the item providers---hence, other people or organizations---are actually impacted and, therefore, the underlying fairness concept aims to convey some sense of social justice related to people.}

In some works, e.g., \cite{BORATTO2021102387}, the \emph{popularity} of items is considered an important attribute. \review{Typical goals in that context are to give fair exposure to items that belong to the long tail, or to include a combination of popular and less popular items in a user-calibrated fashion \citep{Abdollahpouri21User}.}
In other research works that focus on fair item exposure, e.g., in \cite{Gupta2021Online}, groups are defined based on attributes that are in practice not protected \review{in legal terms or based on some accepted normative claim}, e.g., the price range of accommodation.
The purpose of such experiments is usually to demonstrate the effectiveness of an algorithm if (any) groups were given. Nonetheless, in these cases
it often remains unclear in which ways evaluations make sense with datasets from domains where there is no clear motivation for considering questions of fairness. Also, in cases where the goal is to increase the exposure of long-tail items, no particular motivation is usually provided about why recommending (already) popular items is generally unfair. There are often good reasons why certain items are unpopular and should not be recommended, for example, simply because they are of poor quality \citep{DBLP:journals/corr/abs-2109-07946}.

Fairness for items at the \emph{individual} level, in particular for cold-start items, is for example discussed in \cite{Zhu21NewItems}. In general, as shown in Figure~\ref{fig:individual-vs-group-fairness}, works that consider aspects of
individual fairness are
\review{less frequently investigated than group fairness scenarios. An even smaller number of works addresses both types of fairness.}

The definition from classical fair ML settings---similar individuals should be treated similarly---can not always be directly transferred to recommendation scenarios. In \cite{edizel2020fairecsys}, for example, the goal is to make sure that the system is not able to derive a user's sensitive attribute, e.g., gender,
and should thus be able to treat male and female individuals similarly\review{\footnote{It should be noted that if decisions would be based on the protected gender attribute, it would not be individual fairness. In the discussed work, however, the goal is to treat individuals similarly which have similar attributes (and not considering the gender attribute). This then represents an approach towards individual fairness according to the definition.}}.
Most other works that focus on individual fairness address problems of \emph{group recommendation}, i.e., situations where a recommender is used to make item suggestions for a group of users. Group recommendation problems have been studied for many years \citep{Masthoff2011GroupRS,felfernig2018group}, usually with the goal to make item suggestions that are acceptable for all group members and where all group members are treated similarly. In the past, these works were often not explicitly mentioning fairness as a goal, because this was an implicit underlying assumption of the problem setting\review{\footnote{Even though there are some strategies that are not fair, e.g., dictatorship, where one decides for the group \citep{Masthoff2011GroupRS}.}}. In more recent works on group recommendation, in contrast, fairness is
explicitly mentioned, e.g., in \cite{htun2021perception,kaya2020ensuring,Malecek2021Group}, maybe also due to the current interest in this topic. Notable works in this context are \citep{htun2021perception} and \citep{Want2021BiasFriend}, which are one of the few works in our survey which consider questions of fairness \emph{perceptions}.

Finally, we underline the resurgence of the notion of \textit{calibration recommendation} or \textit{calibration fairness} in recommender systems. In ML, calibration is a fundamental concept which occurs when the expected proportions of (predicted) classes match the observed proportions data points in the available data. Similarly, the purpose of calibration fairness is to reflect a measure of the deviation of users' interests from the suggested recommendation in an acceptable proportion \citep{Oh2011,steck2018calibrated,JugovacJannachLerche2017eswa}.
\review{While this may not be inherent and directly related to individual or group fairness, this is the category from this section that better suits such an important (and popular) technique.
In fact, from a conceptual point of view, one may see calibration as implementing a particular form of group fairness, without there being an explicitly protected attribute.
In the entertainment domain, this might be the (implicit) group of independent movie lovers \citep{Abdollahpouri21User};
in the news domain, there may be a group of users who prefer a balanced information offering, e.g., in terms of political opinions.
Applying calibration may then help to avoid that the independent movie lovers receive mainly recommendations of mainstream movies; and that vice versa independent movies obtain a higher chance of exposure.}

\review{More in general, calibration has been applied to either users---by considering age or gender as features to be calibrated against---or items---to compensate for popularity, but also to diversify with respect to item attributes such as genre~\citep{DBLP:journals/ijimai/BobadillaLG021,Abdollahpouri21User,DASILVA2021115112}.
Besides, in works like \citep{Abdollahpouri2020Connection}, calibration is considered as a quality of the recommendations, and the authors measure whether different users or groups experience varying levels of (mis)calibration in their recommendations, since this may indicate an unfair treatment on those populations.
Nonetheless, as stated in \cite{DBLP:conf/ht/LinSMB20}, calibrated recommendations in some domains (such as news or microblogging) might contribute to political polarization in society, so this technique is generally applied to consumer taste domains, where focused, less-diverse recommendations might be valued by users.
Like for other fairness approaches, however, there must be an underlying normative claim that is addressed. Without an underlying normative claim, calibrating recommendations may in some cases merely be a matter of improved personalization and, thus, recommendation quality.}

\paragraph{Single-sided and Multi-Sided Fairness} Traditionally, research in computer science on recommender systems has focused on the consumer value (or utility) of recommender systems, e.g., on how algorithmically generated suggestions may help users deal with information overload. Providers of recommendation services are however primarily interested in the value a recommender can ultimately create for their organization. The organizational impact of recommender systems has been, for many years, the focus in the field of information systems, see \citep{Xiao:2007:EPR:2017327.2017335} for a survey. Only in recent years we observe an increased interest on such topics in the computer science literature. Many of these recent works aim to shed light on the impact of recommendations in a multistakeholder environment, where typical stakeholders may include consumers, service providers, suppliers of the recommendable items, or even society \citep{abdollahpouri2020,jannach2021mcnamara}.

In multistakeholder environments, there may exist trade-offs between the goals of the involved entities. A recommendation that is good for the consumer might for example not be the best for the profit perspective of the provider \citep{JannachAdomaviciusVAMS2017}. In a similar vein, questions of fairness can be viewed from multiple stakeholders, leading to the concept of \emph{multisided} fairness \citep{Burke2017Fairness}, \review{ which might include the utility of system designer and other side-stakeholders in addition to the consumer and provider.}
As mentioned above, there can be fairness questions that are related to the providers of the items.
Again, there can also be tradeoffs \review{and in some ways incompatible notions of fairness}, i.e., what may be a fair recommendation for users might be in some ways be seen to be unfair to item providers, e.g., when their items get limited exposure \review{\citep{DBLP:journals/corr/abs-2009-02423}}.

\begin{figure*}[h!tb]
     \centering
     \includegraphics[width=0.5\linewidth]{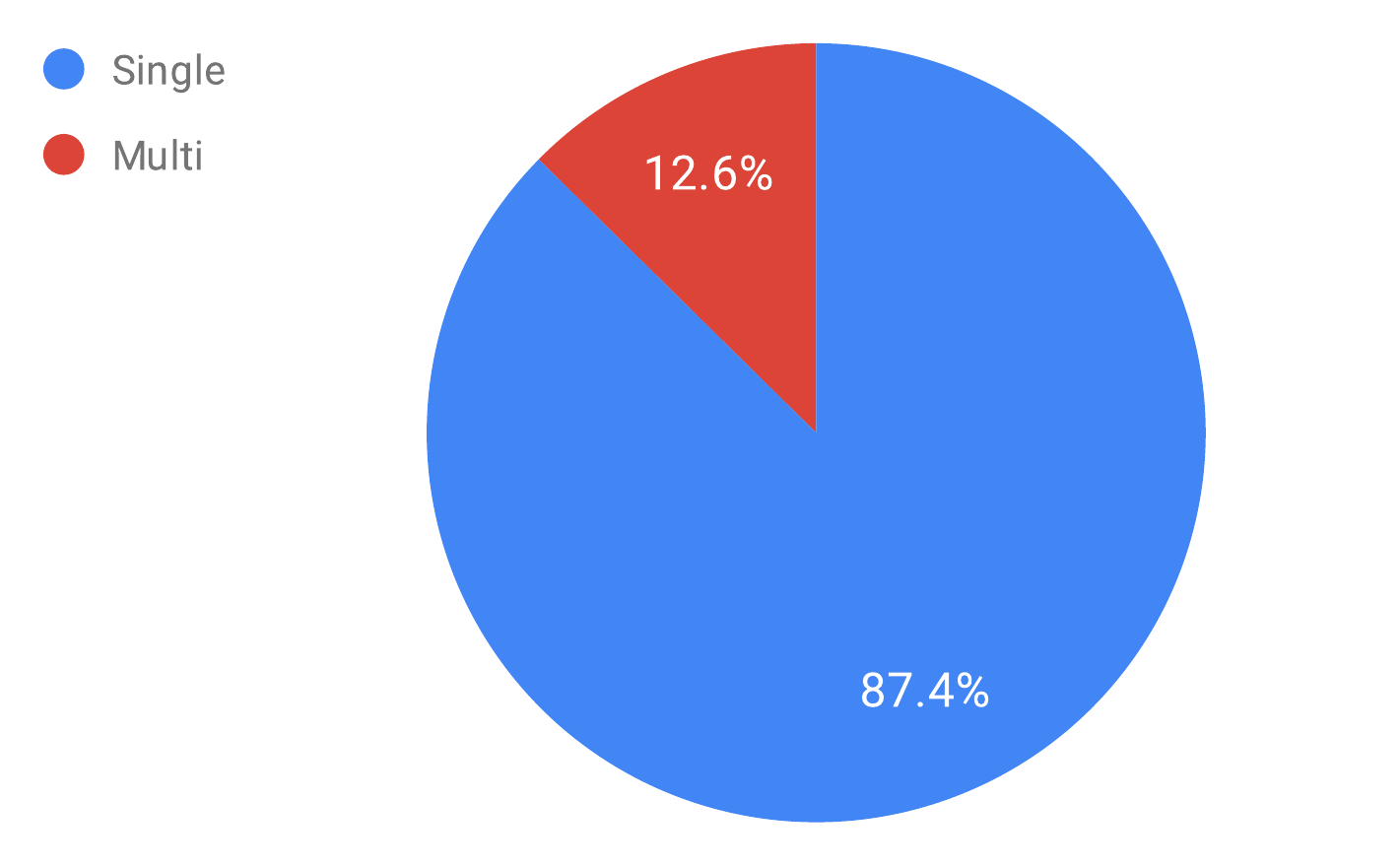}
     \caption{Fairness Notions: Single-sided vs.~Multi-sided Fairness.
     }
     \label{fig:single-sided-vs-multi-sided}
 \end{figure*}

Figure~\ref{fig:single-sided-vs-multi-sided} shows the distribution of works that focus on one single side of fairness and works which address questions of multisided fairness. The illustration clearly shows that the large majority of the works concentrates on the single-sided case, indicating an important \emph{research gap} in the area of multisided fairness within multistakeholder application scenarios.

Among the few studies on multi-sided fairness, \citep{Abdollahpouri19Multi}
discusses techniques for CP-fairness in matching platforms such as Airbnb and Uber. In~\cite{rahmani2022role}, the authors explore how adding contextual information such as geographical, temporal, social, and categorical affects the multi-aspect quality of POI suggestions, including accuracy, beyond-accuracy, fairness, and interpretability (see also \citep{rahmani2022exploring} for a discussion on a temporal bias).~\citep{patro2020fairrec} model the fair recommendation problem as a constrained fair allocation problem with indivisible goods and propose a recommendation algorithm that takes producer fairness into consideration. In \cite{anelli2023auditing} the authors study  the CP-Fairness in several graph CF models. \cite{Wu21TFROM}
propose an individual-based perspective, where fairness is defined as the
same exposure
for all producers and the same NDCG for all consumers involved.
\review{Exposure in this work is defined based on the appearance of items of providers on top-n recommendation lists, where a higher ranking is assumed to lead to higher exposure.}

\paragraph{Static vs.~Dynamic Fairness}
Another dimension of fairness research relates to the question whether the fairness assessment is done in a static or dynamic environment \citep{Li2021tutorial}. In static settings, the assessment is done at a single point of time, as commonly done also in offline evaluations that focus on accuracy. Thus, it is assumed that the attributes of the items do not change, that the set of available items does not change, and that the analysis that is made at one point in time is sufficient to assess the fairness of algorithms or if an unfairness mitigation technique is effective.

Such static evaluations however have their shortcomings, e.g., as there may be feedback loops that are induced by the recommendations. Also, some effects of unfairness and the effects of corresponding mitigation strategies might only become visible over time. Such longitudinal studies require alternative evaluation methodologies, for example, approaches based on synthetic data or different types of \emph{simulation}, such as those developed in the context of reinforcement learning algorithms, see \citep{rohde2018recogym,mladenov2021recsimng,ghanem2022balancing,longitudinalimpact2021,AdomaviciusJannach2021} for simulation studies and
related frameworks in recommender systems.

  \begin{figure*}[h!tb]
     \centering
     \includegraphics[width=0.5\linewidth]{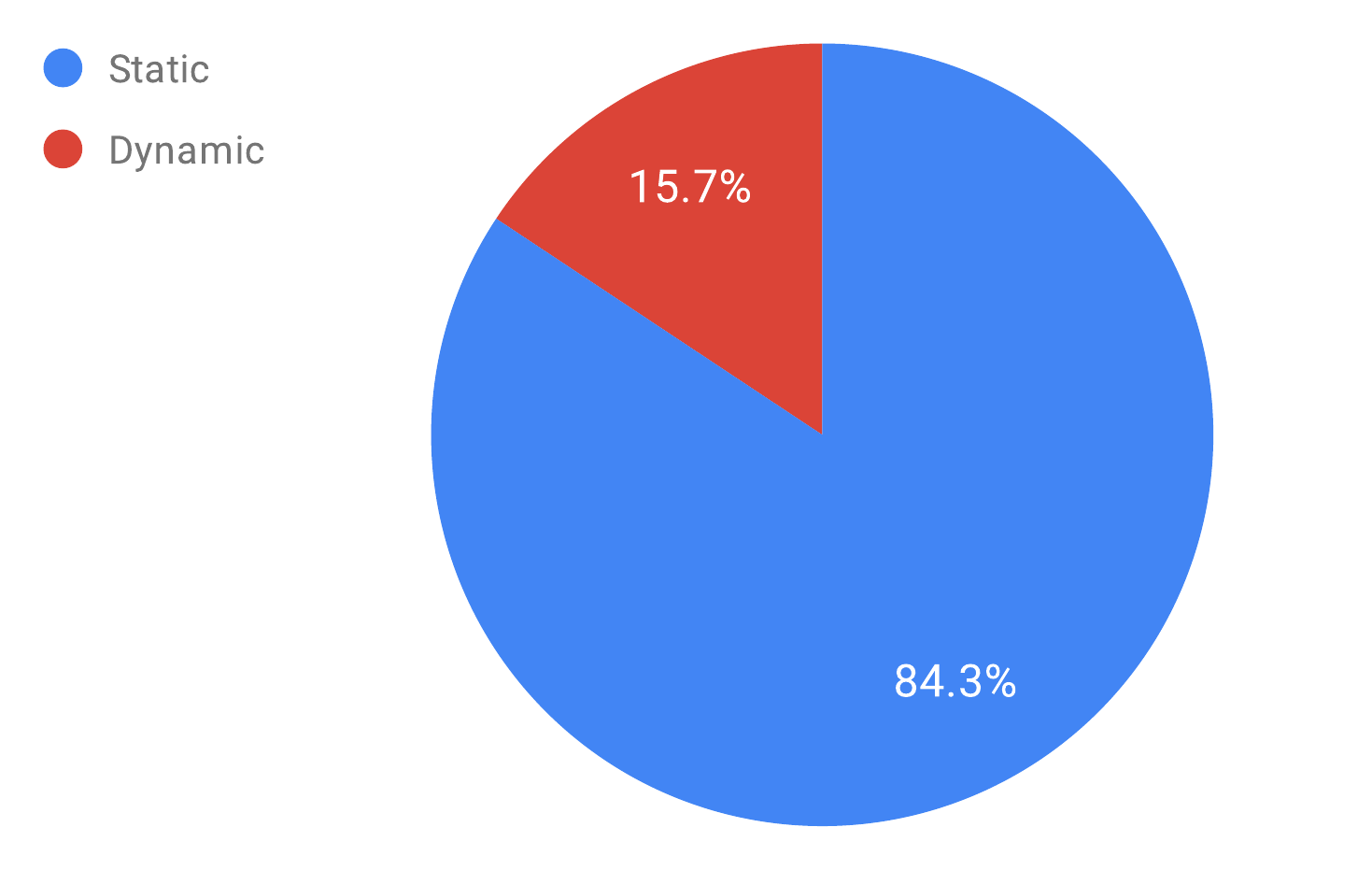}
     \caption{Fairness Notions: Static vs.~Dynamic Fairness Evaluation}
     \label{fig:static-vs-dynamic-evaluation}
 \end{figure*}

Figure~\ref{fig:static-vs-dynamic-evaluation} shows how many studies in our survey considered static and dynamic evaluation settings, respectively. Static evaluations are clearly predominant: we only found \review{16} works that consider dynamically changing environments.
In \cite{Ge21Towards}, for example, the authors consider the dynamic nature of the recommendation environment by proposing a fairness-constrained reinforcement learning algorithm so that the model dynamically adjusts its recommendation policy to ensure the fairness requirement is satisfied even when the environment changes.
A similar idea is developed in \cite{DBLP:conf/pakdd/LiuLTLCH20}, where a long-term balance between fairness and accuracy is considered for interactive recommender systems, by incorporating fairness into the reward function of the reinforcement algorithm.
\review{Moreover, in \cite{DBLP:journals/corr/abs-2009-02590}, a framework is proposed for the dynamic adaptation of recommendation fairness using \emph{Social Choice}. The goal of this work is to arbitrate between different re-ranking methods, aiming to achieve a better accuracy-fairness tradeoff with respect to all sensitive features.}
On the other hand, works such as \citep{DBLP:conf/kdd/BeutelCDQWWHZHC19} and \citep{deldjoo2021flexible}
model fairness in a specific snapshot of the system, by simply taking the system and its training information as a fixed image of the interactions performed by the users on the system.

\paragraph{Associative vs.~Causal Fairness}
The final categorization discussed in \cite{Li2021tutorial} contrasts \emph{associative} and \emph{causal} fairness. One key observation by the authors in that context is that most research in fair ML is based on association-based (correlation-based) approaches. In such approaches, researchers typically investigate the potential \emph{``discrepancy of statistical metrics between individuals or subpopulations}''. However, certain aspects of fairness cannot be investigated properly without considering potential causal relations, e.g., between a sensitive (protected) feature like gender and the model's output. In terms of methodology, causal effects are often investigated based on counterfactual reasoning \citep{KusnerCounterfactual2017,li-2021-towards-1}.

\begin{figure*}[h!tb]
     \centering
     \includegraphics[width=0.5\linewidth]{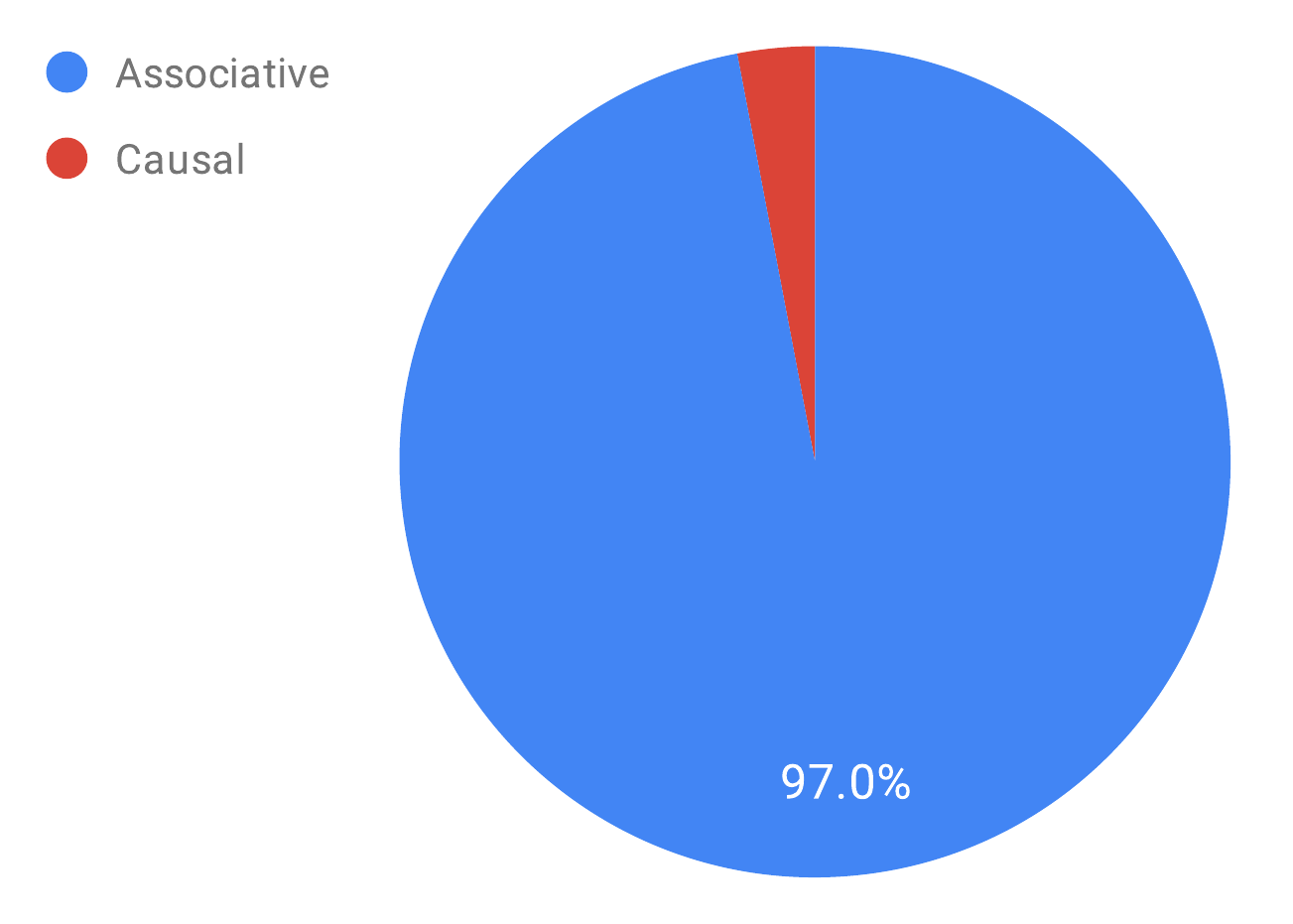}
     \caption{Fairness Notions: Associative vs.~Causal Fairness.}
     \label{fig:associative-vs-causal-fairness}
 \end{figure*}

Figure~\ref{fig:associative-vs-causal-fairness} shows that there are
only \emph{three} works
investigating recommendation fairness problems based on causality considerations.
\review{More specifically,}
in \cite{DBLP:conf/recsys/CornacchiaNR21},
the authors propose the use of counterfactual explanation to provide fair recommendations in the financial domain.
An interesting alternative is presented in \cite{li-2021-towards-1}, where the authors analyze the causal relations between the protected attributes and the obtained results.
\review{The third work we found in our review, \cite{DBLP:conf/mm/QiuWCYH21}, derives a causal graph to identify and analyze the visual bias of existing methods, so that spurious relationships between users and items can be removed.}

One additional dimension we have discovered through our literature analysis is the use of \review{\emph{constraint-based approaches}} to integrate or model fairness characteristics in recommender systems.
\review{In this context, these approaches may be seen as an alternative paradigm to associative and causal inference, which is based on explicit constraints and special techniques, often from multi-objective optimization, to achieve the desired fairness goals.}
For example, \cite{Hao21Pareto} address the issue of enforcing equality to biased data by formulating a constrained multi-objective optimization problem to ensure that sampling from imbalanced sub-groups does not affect gradient-based learning algorithms; the same work and others---including \citep{DBLP:conf/recsys/SeymenAM21a} or \citep{DBLP:conf/sigir/YadavDJ21}---define fairness as another constraint to be optimized by the algorithms. In \cite{DBLP:conf/sigir/YadavDJ21}, in particular, such a constraint is amortized fairness-of-exposure.

\subsection{Application Domains and Datasets}\label{ss:domains}
\begin{figure*}[h!tb]
     \centering
     \includegraphics[width=0.9\linewidth]{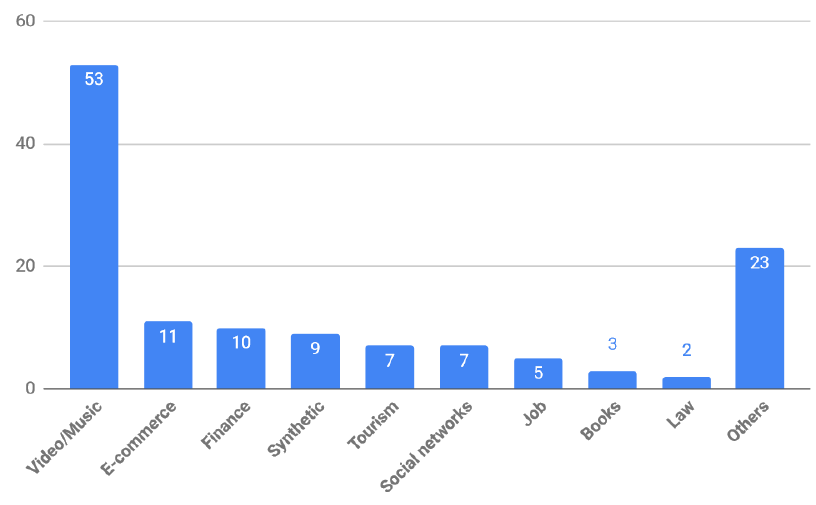}
     \caption{\review{Application domains of used datasets. Note that some studies rely on more than one dataset, and a number of theoretical or conceptual works do not provide experimental validation.}}
     \label{fig:application-domains}
 \end{figure*}

Next, we look at application domains that are in the focus of research on fair recommendations.  Figure~\ref{fig:application-domains} shows an overview of the most frequent application domains and how many papers focused on these domains in their evaluations.\footnote{\review{The categorization of the papers was based on the datasets that were used for the empirical evaluations. We used higher-level categories of domains as done in earlier surveys, e.g., in \cite{NunesJannachUmuai2017,JannachZankerEtAl2012}.}} The by far most researched domain is the recommendation of videos (movies) and music, followed by \review{e-commerce, and finance.} For many other domains shown in the figure (e.g., jobs, tourism, or books), only a few papers were identified. Certain domains were only considered in one or two papers. These papers are combined in the ``Other'' domain in Figure~\ref{fig:application-domains}.

Since most of the studied papers are technical papers and use an offline experimental procedure, corresponding datasets from the respective domains are used. Strikingly often, in more than one third of the papers, one of the MovieLens datasets is used. This may seem surprising as some of these datasets not even contain information about sensitive attributes. Generally, these observations reflect a common pattern in recommender systems research, which is largely driven by the availability of datasets. The MovieLens datasets are a \review{widely adopted and probably overused}
case and have been used for all sorts of research in the past \citep{ML2015}. Fairness research in recommender systems thus seems to have a quite different focus than fair ML research in general, which is often about avoiding discrimination of people.

We may now wonder which specific fairness problems are studied with the help of the MovieLens rating datasets.
What would be unfair recommendations to users? What would be unfair towards the movies (or their providers)? It turns out that item popularity is often the decisive attribute to achieve \emph{fairness towards items}, and quite a number of works aim to increase the exposure of long-tail items which are not too popular, see, e.g., \cite{DBLP:journals/access/DongXL21}. In terms of \emph{fairness towards users}, the technical proposal in \cite{DASILVA2021115112} for example aims to serve users with recommendations that reflect their past diversity preferences with respect to movie genres. An approach towards
\emph{\review{group fairness}}
is proposed in \cite{MISZTALRADECKA2021102519}. Here, groups are not identified by their protected attribute, but by the recommendation accuracy that is achieved (using any metric) for the members of the group.

\review{In other domains beyond Video/Music (dominated, as mentioned above, by MovieLens datasets), fairness is characterized by the inherent properties of users and items in each particular domain. For example, in e-commerce the price or year of the item, or the helpfulness of the provided user's review are considered \citep{deldjoo2021flexible}; in tourism, the user's gender and the business category are typically analyzed \citep{DBLP:conf/recsys/MansouryMBP19}.}

Continuing our discussions above, such notions of unfairness \review{in the described application contexts} may not be undisputed. When some users receive recommendations with lower accuracy, this might be caused by their limited activity on the platform or their unwillingness to allow the system to collect data. Actually, one may consider it unfair to artificially lower the quality of recommendations for the group of highly active and open users. \review{In another example, it might not be clear why recommending \review{less popular items}---which might in fact not be popular because of their limited quality---}would make a system fairer, and equating bias (or skewed distributions) with unfairness in general seems questionable. \review{Therefore, we iterate the importance of clearly specifying the underlying assumptions, hypothesis, and normative claims in any given research work on fairness. Otherwise it may remain unclear to what extent a particular system design or algorithmic approach will ensure or increase a system's level of fairness.}

\review{Similar questions arise when using calibration approaches to ensure fairness in a personalized, user-individual way.}
\review{Considering, for example, a} user fairness calibration approach \review{like the one presented in} \cite{DASILVA2021115112}, it is less than clear why diversifying recommendations according to user tastes would increase the system's fairness. It may increase the quality of the recommendations, but a system that generates recommendations \review{of limited quality in terms of calibration} for everyone is probably not one we would call unfair.
\review{However, note that there actually \emph{may be} situations where calibration serve a certain fairness goal. Consider, for example, that a recommendation provider notices that users with niche tastes often receive item recommendations that are not interesting to them. This may happen when an algorithm too strongly focuses on mainstream items and when the used metrics do not reveal clearly that there are some user groups that are not served well. Under the assumption that users with niche tastes might also be users who are marginalized in other ways, e.g., when they are users who differ because of ethnicity or national origin, then improving calibration may indeed serve a fairness goal. These assumptions and claims however have to be made explicit, as otherwise it might just be an issue of whether the recommendation quality is measured in the right way.}

In several cases, \review{and independent of the particular application domain}, it therefore seems that the addressed problem settings are not too realistic or \review{remain artificial to a certain extent.} One main reason for this phenomenon in our view lies in the lack of suitable datasets for domains where fairness really matters. These could for example be the problem of job recommendations on business networks or people recommendations on social media which can be discriminatory. In today's research, often datasets from rather non-critical domains or synthetic datasets are used to showcase the effectiveness of a technical solution \citep{Ge21Towards,Abdollahpouri21User,Yao2017BeyondParity,MISZTALRADECKA2021102519,Hao21Pareto,Tsintzou19BiasDisparity,Sun19Debiasing,DBLP:conf/kdd/GeyikAK19,Stratigi17health}.
 While this may certainly be meaningful to demonstrate the effects of, e.g., a fairness-aware re-ranking algorithm, such research may appear to remain quite disconnected from real-world problems.
\review{Related phenomena of ``abstraction traps'' in fair ML were discussed earlier in \cite{selbst2019}. While abstraction certainly is central to computer science, the danger exists that central domain-specific or application-specific idiosyncrasies are abstracted away so that ML tools can be applied. In the end, the proposed solutions for the abstracted problem may then fail to properly account for the sometimes complex interactions between technical systems and the real world, and to respond to the \emph{``fundamental tensions, uncertainties, and conflicts inherent in sociotechnical systems.''}~\citep{selbst2019}}

\subsection{Methodology} \label{ss:meth}
In this section, we review how researchers approach the problems from a methodological perspective.

\begin{figure*}[h!tb]
     \centering
     \includegraphics[width=0.6\linewidth]{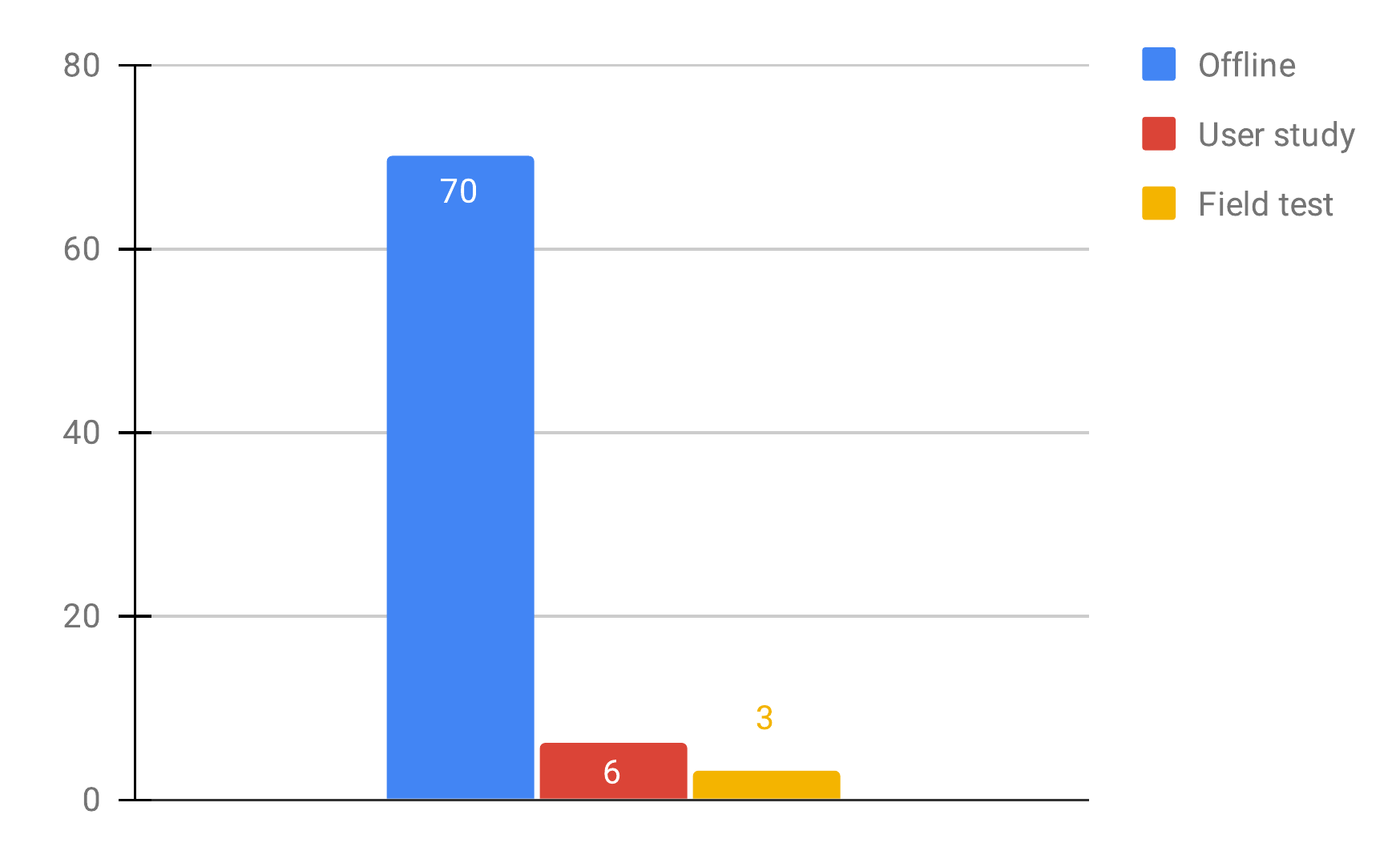}
     \caption{Experiment Types.}
     \label{fig:experiment-types}
 \end{figure*}

\paragraph{Research Methods}
In principle, research in recommender systems can be done through experimental research (e.g., with a field study or through a simulation) or non-experimental research (e.g., through observational studies or with qualitative methods) \citep{reference/rsh/ShaniG11,JannachZankerEtAl2010}. In recommender systems research, three main types of experimental research are common: (a) offline experiments based on historical data, (b) user studies (laboratory studies), and (c) field tests (A/B tests, where different systems versions are evaluated in the real world). Figure~\ref{fig:experiment-types} shows how many papers fall into each category. Like in general recommender systems research \citep{JannachZankerEtAl2012}, we find that offline experiments are the predominant form of research.
Note that we here only consider \review{83} technical papers, and not the conceptual, theoretical, and analytic ones that we identified. Only in very few cases (\review{6 papers}), humans were involved in the experiments, and in even fewer cases (\review{3 papers}) we found reports of field tests. Regarding user studies, \cite{htun2021perception} for example involves real users to evaluate fairness in a group recommendation setting. On the other hand, notable examples of field experiment are provided in \cite{DBLP:conf/kdd/GeyikAK19}, where a gender-representative re-ranker is deployed for a randomly chosen 50\% of the recruiters on the \textit{LinkedIn Recruiter} platform (A/B testing), and
in \cite{DBLP:conf/kdd/BeutelCDQWWHZHC19}\review{, where the engagement with a large-scale recommender system in production is reported across sub-groups of users}.
We only found one paper that relied on interviews as a qualitative research method \citep{Sonboli2021Fairness}. Also, only very few papers used more than one experiment type, e.g., \cite{Serbos17package} were both a user study and an offline experiment were conducted.

The dominance of offline experiments points to a research gap in terms of our understanding of \emph{fairness perceptions} by users. Many technical papers that use offline experiments assume that there is some target distribution or a target constraint that should be met. And these papers then use computational metrics to assess to what extent an algorithm is able to meet those targets. The target distribution, e.g., of popular and long-tail content, is usually assumed to be given or to be a system parameter. To what extent a certain distribution or metric value would be considered fair by users or other stakeholders in a given domain is usually not discussed. In any practical application, this question is however fundamental, and
again the danger exists that research is stuck in an abstraction trap, \review{as characterized above}. In a recent work on job recommendations \citep{Want2021BiasFriend}, it was for example found that a debiasing algorithm lead to fairer recommendation without a loss in accuracy. A user study then however revealed that participants actually preferred the original system recommendations.

\begin{figure*}[h!t]
     \centering
     \includegraphics[width=0.6\linewidth]{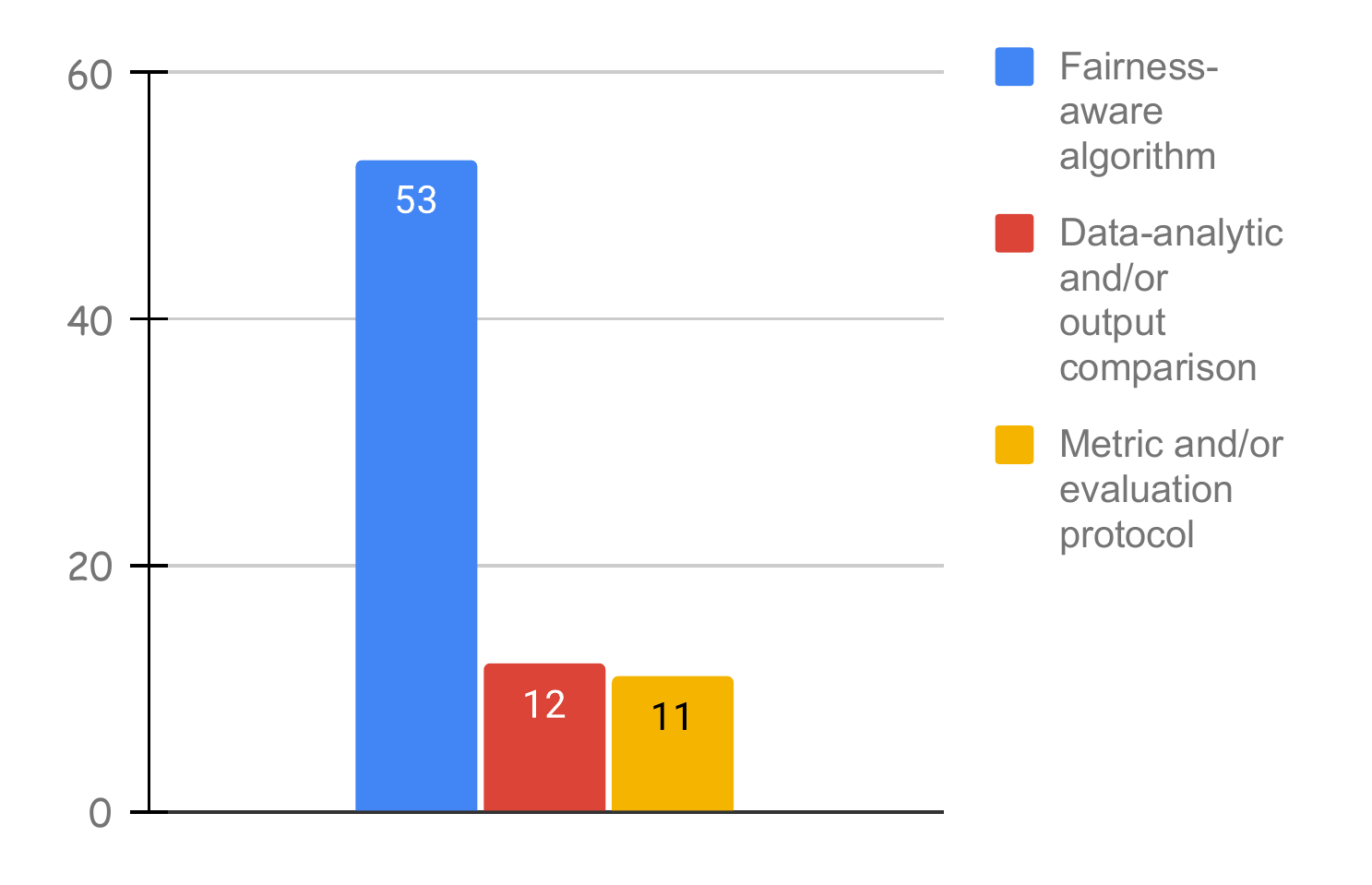}
     \caption{Technical Focus of Papers.}
     \label{fig:technical-focus}
 \end{figure*}

\paragraph{Main Technical Contributions and Algorithmic Approaches}
Looking only at the \emph{technical} papers, we identified three main groups of technical contributions: (i) works that report outcomes of data analyses or which compare recommendation outcomes, (ii) works that propose algorithmic approaches to increase the fairness of the recommendations, and (iii) works that propose new metrics or evaluation approaches. Figure~\ref{fig:technical-focus} shows the distribution of papers according to this categorization.

\begin{figure*}[h!tb]
     \centering
     \includegraphics[width=0.6\linewidth]{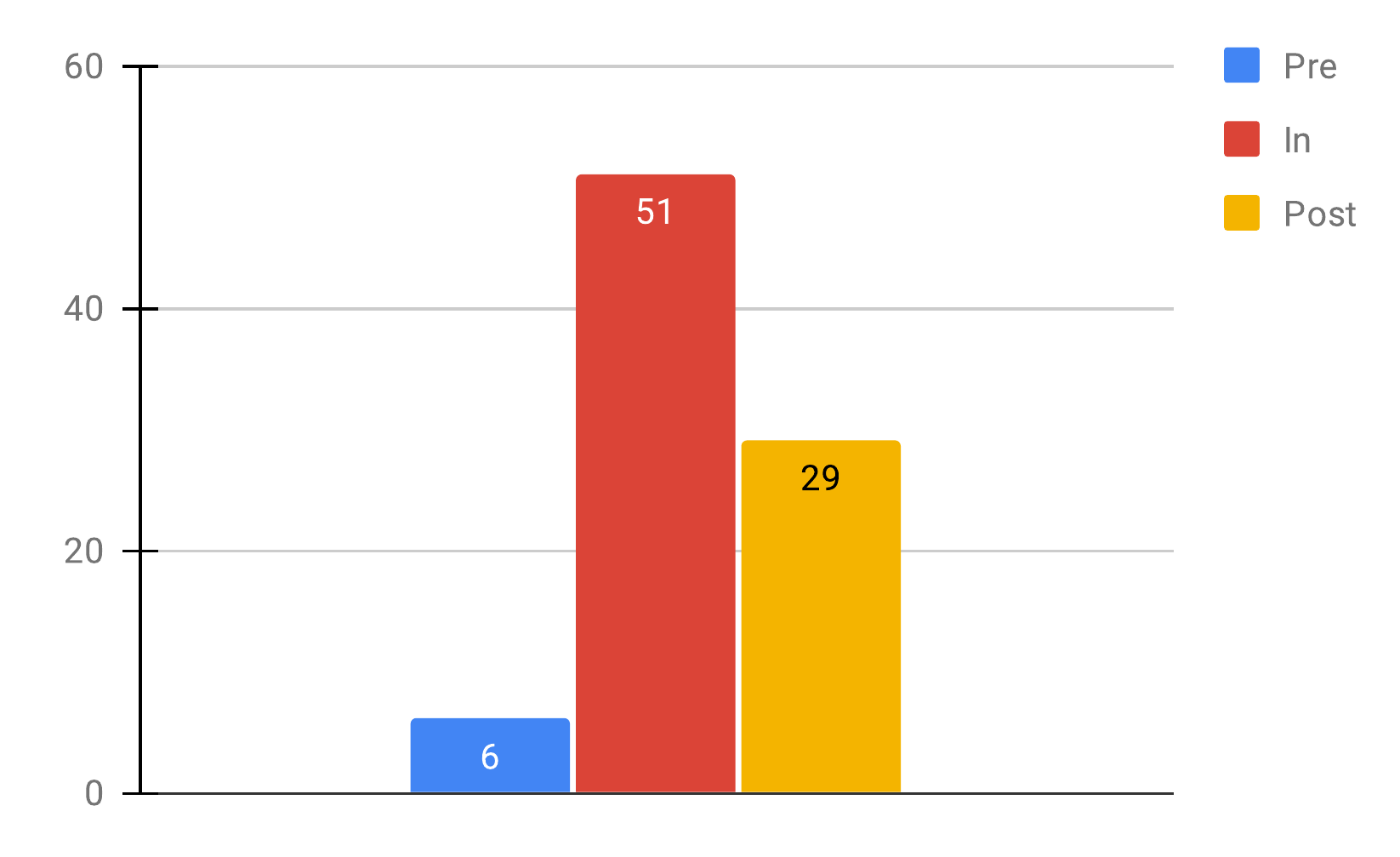}
     \caption{Fairness Step.}
     \label{fig:fairness-step}
 \end{figure*}

We observe that most technical papers aim to make the recommendations of a system fairer, e.g., by reducing biases or by aiming to meet a target distribution. Technically, in analogy to context-aware recommender systems \citep{Adomavicius2015}, this ``fairness step'' can be done (i) in a pre-processing step, (ii) integrated in the ranking model (modeling approaches), or (iii) in a post-processing step. Figure~\ref{fig:fairness-step} shows what is common in the current literature, \review{see also \citep{Li2022Survey}}. Methods that rely on some form of pre-processing are comparably rare.
Typical approaches for modeling approaches include specific fairness-aware loss functions or optimizing methods that consider certain constraints. Post-processing approaches are frequently based on re-ranking.

Overall, the statistics on the one hand point to a possible research gap in terms of works that aim to understanding what leads to unfair recommendations and how severe the problems are for different algorithmic approaches in particular domains. In the future, it might therefore be important to focus more on analytical research, as advocated also in \cite{jannach2021mcnamara}, e.g., to understand the idiosyncrasies of a particular application scenario instead of aiming solely for general-purpose algorithms. On the other hand, the relatively large amount of work that propose new ways of evaluating indicate that the field is not yet mature and has not yet established a standardized research methodology. We discuss evaluation metrics next.

\review{\paragraph{Evaluation Metrics.} In offline experiments, a variety of computational metrics
are employed to evaluate the fairness of a set of recommendations. The choice of a certain fairness metric is mostly determined by the underlying concept of fairness, such as whether it is about individual or group fairness. In Table~\ref{table:group-fairness-metrics} and Table~\ref{table:individual-fairness-metrics}, we provide detailed lists of selected metrics used in the literature on fairness in recommender systems.\footnote{\review{We note that in these tables we only provide individual examples of works that used a particular metric.}} We primarily organize the metrics along the common categorization of group fairness (Table~\ref{table:group-fairness-metrics}) vs.~individual fairness (Table~\ref{table:individual-fairness-metrics}).
Within the category of \textit{group fairness} metrics, we furthermore mainly distinguish between the types of \textit{utility} (benefit) in terms of \emph{exposure} and \emph{effectiveness}~\citep{amigo2022unifying}. The metrics listed in Table~\ref{table:individual-fairness-metrics}, in contrast, are split into \emph{(a)} metrics for  individual item recommendation scenarios, and \emph{(b)} metrics for \emph{group recommendation} settings.}
\review{ Exposure and effectiveness can be defined as follows:
\begin{itemize}
    \item \textit{Exposure} refers to the degree to which an item or group of items is exposed to a user or group of users;
    \item \textit{Effectiveness} (sometimes called \emph{relevance}) defines the amount to which an item's exposure is effective, i.e., corresponds to the user's preferences. \vspace{1mm}
\end{itemize}
Different stakeholders in recommender systems may be concerned with these two types of utility to varying degrees. For instance, from the perspective of customers, fairness primarily entails an equitable distribution of effectiveness among users, thereby preventing the discrimination of historically disadvantaged groups such as female or black job applicants, for example. In contrast, producers and item providers that seek enhanced visibility are primarily concerned with exposure equity, which should not be punished, for instance, based on producers' popularity or country.}

\review{We note that the popularity of items is a central concept in most metrics that are related to \emph{exposure}. Most commonly, the popularity of an item is assessed in offline experiments by counting the number of observed interactions for each item in the training data.}
\review{Moreover, various work assume that there is a trade-off between different evaluation objectives: customer fairness, provider fairness, and overall system accuracy. Thus, some metrics in the literature are designed against the background of such potential trade-offs.}

\footnotesize
\color{black}
\begin{longtable}{@{ }p{3.5cm} p{7.6cm}@{ }}
\caption{Selected types of evaluation metrics used for \emph{group fairness} scenarios.}
\label{table:group-fairness-metrics}\leavevmode\\
\toprule
    \multicolumn{2}{l}{\bf Metrics used for measuring Exposure}\leavevmode\\ 
    \midrule
        Popularity of recommended items  & Different measures are used in the literature to quantify the popularity of the items in a given list of recommendations, e.g., the Average Recommendation Popularity (ARP) \cite{Abdollahpouri19Managing} or the PCOUNT measure in~\cite{Borges21mitigating}. The assumption is that recommending less popular items increases fairness, see also \cite{deldjoo2021explaining}.\\  

        Deviation from popularity-ranked list & In~\cite{Borges21mitigating}, the authors propose a metric to assess the popularity bias of a list inspired by the Normalized Cumulative Gain (NDCG) metric. The popularity bias is assessed by comparing a given top-n recommendation list with a list that is ranked by popularity. Lists which differ more strongly from a pure popularity-ranked  list are considered to be fairer. \\
        Proportion of less popular items in recommendations & Different metrics in the literature assess the number of less popular (long-tail) items in the top-n recommendations as a fairness indicator. These metrics are called Average Percentage of Long Tail Items (APLT) in \cite{Abdollahpouri19Managing} or Popularity Rate in~\cite{Ge21Towards}. Such metrics are commonly based on some pre-defined threshold to distinguish long-tail items from other. \leavevmode\\
        Disparate exposure of provider groups & In~\cite{Boratto21Interplay}, the authors compare how often the items of a certain group of item providers are recommended relative to the proportion of items of this provider group in the catalog. This measure is used to assess what the authors term ``disparate visibility''. A variation of this measure, ``disparate exposure'', also includes a positional decay, see also~\citep{GomezWinner2021}. The underlying fairness assumption is that items of a minority group of providers should be recommended to users proportional to their representation.  \\
        Individual provider exposure & Different exposure-based metrics were proposed which assume that items from the same provider belong to the same group. In~\cite{Wu21TFROM}, the variance of the distribution of group-level exposures is used, whereas in~\cite{patro2020fairrec} an entropy-like measure is used; in both cases, a lower value evidences less inequality and, hence, more fairness. In~\cite{patro2020fairrec} another metric is defined based on a minimum exposure requirement (i.e., each product must be assigned to a minimum number of distinct customers) to measure the fraction of \textit{satisfied} producers. \\

        Variance of provider exposure & Also~\cite{Wu21TFROM} base their fairness assessments on the exposure of the items of providers relative to the number (and quality) of their items in the catalog (as in \cite{Boratto21Interplay}). The final fairness judgment for a recommender system is however then made by considering the \emph{variance} of exposures across providers (groups), where lower variance indicates higher fairness. \\
        
        Ranking-based Statistical Parity (RSP) & \cite{Zhu2020Measuring} propose to assess if items of different provider groups have the same probability to be contained in the top-$k$ recommendation lists of users. A system is considered fair if it ensures statistical parity, i.e., when the probability distributions of being ranked (exposed in) in top-$k$ lists are comparable for different groups. \\

        Divergence of exposure probabilities &  In~\cite{Dash21umpire}, the authors aim to assess the probability of exposure for ``sponsored'' recommendations compared to ``organic'' recommendations on e-commerce marketplaces. To that purpose they compute the Kullback-Leibler divergence of the distributions, which they estimate based on different factors. A system is considered fair and not exhibiting exposure bias when the divergence is close to zero. \\

        Concentration on a subset of items & A number of works, e.g., \cite{Ge21Towards}, use the Gini index to assess to what extent a recommender system has a tendency to focus on a limited set of items. Such a concentration on a subset of the items in the catalog may lead to an overproportional, and thus unfair, exposure of some items. The Gini index is a number between 0 and 1, which is traditionally used to quantify inequalities, e.g., in terms of income in a society. A higher Gini index means higher concentration. We however note that this not necessarily means that the concentration is on popular items (which is however usually the case in practice). \\

    \midrule
    \multicolumn{2}{l}{\bf Metrics used for measuring Effectiveness\footnote{\review{Historically, evaluations of fairness in recommender systems mostly associated \dquotes{exposure} with \dquotes{providers} and \dquotes{effectiveness} with \dquotes{consumers}, as these utilities are of most interest to these stakeholders. However, other works use less explored scenarios, e.g., \citep{Boratto21Interplay,Zhu2020Measuring}, and examine effectiveness from the provider perspective.}}}\leavevmode\\ 
    \midrule Difference between group's utility & The simplest way to evaluate group fairness is to calculate the \textit{difference} (typically in an absolute sense) in the \textit{average} performance of group members where groups are defined based on the protected attributes; here the performance can be quantified using ranking-aware (e.g., NDCG), or rating-based measures (e.g., RMSE). This concept is used to quantify group fairness in a number of publications under several titles, including mean Absolute Difference~\cite{DBLP:conf/cikm/ZhuHC18,deldjoo2021flexible,deldjoo2021explaining}, or user-oriented group fairness (UGF)~\cite{Li2012UserOriented}, and even Negative bias~\cite{MISZTALRADECKA2021102519}, where the latter calculates the difference between a performance metric (e.g., NDCG) for a user segment and all other users. It should be highlighted that this metric can be utilized to measure producers' exposure fairness, see e.g.,~\cite{deldjoo2021flexible}.\\
    Relevance disparity
    &  \review{This metric was introduced along with \dquotes{Disparate exposure of provider groups} from above in~\cite{Boratto21Interplay}. Essentially, this paper examines the same disparity on the producer-side, but with relevance as the underlying utility. The authors note that a disparity in relevance values might not necessarily imply that the minority group is discriminated against based on its exposure or visibility in the recommendations lists, but it may be exacerbated through continuous recommendation loops.}
   \\
        Prediction error access market segment
        &
        The average prediction errors of a fair algorithm are supposed to be similar for different market segments. Thus, in \cite{wan2020addressing} the authors propose to use statistical significance tests and the F-statistic as a fairness evaluation metric to evaluate a global parity of prediction errors across different consumer-product market segments. Lower values in this approach indicate better rating prediction fairness. 
        \vspace{1.5mm}\leavevmode\\
        Ranking-based equal opportunity (REO) & This metric, again introduced by~\cite{Zhu2020Measuring}, is similar to RSP presented in the previous Table but is primarily concerned with measuring effectiveness fairness. It quantifies the discrepancy between item groups based on the probability that a relevant item is among the top-$k$ suggestions; \\
        \midrule
    \multicolumn{2}{l}{\bf Other Scenarios}\\
    \midrule
  & \\
Two-sided metrics & A number of metrics were proposed that integrate two group fairness criteria, namely consumer effectiveness and consumer exposure.
(i) \textit{Flexible probabilistic metrics:} Some works have presented fairness measurement models that are adaptable to specific scenarios,
mostly by comparing the distributions provided by a given system against an ideal (fair) distribution,
sometimes called \textit{target representation}, see~\citep{Kirnap21Estimation,amigo2022unifying}. Generalized Cross Entropy~\citep{deldjoo2021flexible,Deldjoo19GCE,rahmani2022role} 
is such a metric
that compares 
those
two distributions.
Similarly,~\cite{Kirnap21Estimation} investigate a variety of divergence-based metrics and target representation types (e.g., based on equity, proportionality to the corpus size, etc.);
(ii) \textit{Joint multi-sided metrics:} another group of fairness metrics eliminates the constraint of comparing against a target representation and evaluates fairness on the basis of statistical independence between user and item groups. Examples include Bias Disparity~\citep{Tsintzou19BiasDisparity,DBLP:conf/recsys/MansouryMBP19, DBLP:conf/recsys/LinSMB19} and Mutual Information~\citep{amigo2022unifying}.
Another example is \cite{Wu21TFROM}, where the authors study joint multi-sided fairness evaluation by designing  metrics to measure the individual fairness of customers, group fairness of providers, and the overall quality of the recommendation results by measuring the \textit{quality-weighted exposure} for the provider side and comparing the reduction in individuals' recommendation quality for the consumer side (see individual fairness).

\\

Calibration  &
The assumption behind calibration metrics is that fair recommendations should not deviate from the historical data of the user, this is exactly what User Popularity Deviation (UPD)~\citep{Abdollahpouri21User} measures in terms of the user's interest towards popular items. $\Delta GAP$ (Group Average Popularity) by \cite{Wundervald21Cluster} measures the same, but at the (user) group level;
\\
        Weighted Proportional Fairness &
        Inspired by rate control algorithms for communication networks, this metric proposed in~\cite{DBLP:conf/pakdd/LiuLTLCH20} is a generalized Nash solution that seeks equilibrium when allocating items (associated with a category) to users. For this, it solves a constrained maximization problem based on the exposure of each group of items.
        \\

    \bottomrule
\end{longtable}
\color{black}
\normalsize

\footnotesize
\color{black}
\begin{longtable}{@{ }p{3.5cm} p{7.6cm}@{ }}
\caption{Selected types of evaluation metrics used for \emph{individual fairness} scenarios.}
\label{table:individual-fairness-metrics}\leavevmode\\
    \toprule
    \multicolumn{2}{l}{\bf Individual recommendation scenario}\leavevmode\\
    \midrule
        Max individual deviation & Addressing the potential trade-off between fairness and other domain-specific requirements/utilities is important. For instance, in particular applications such as mobile apps for video recommendation, regulating fairness, and improving network gains are both crucial goals that may be at odds. Thus, \cite{Giannakas21network} study the problem of network-friendly recommendation (NFR) focusing on owner/producer satisfaction, as measured by the difference in exposure opportunity provided to a single piece of content (item) between the fair recommender being evaluated and a baseline NFR. Individual fairness depends on the metric calculated based on individual content disparity not exceeding a maximum threshold (worst-case scenario),  as indicated by the maximum individual deviation. The authors also apply other aggregation metrics, such as total variation distance and Kullback-Leibler Divergence, which eliminate the constraint for individual content and instead concentrate on the disparity on the provider level (group fairness).              \vspace{1.5mm} \leavevmode\\

        The variance of individual losses & In certain research studies, the same trade-off is handled by assuming that the quality of recommendations will decrease when providers' fair exposure is taken into account, and more importantly, that individual fairness can be measured by \textit{reduction} of individual user recommendation quality. Therefore, it is possible to define individual unfairness as the differences in user losses and to seek for this individual loss value to be dispersed evenly to each consumer, as measured by the difference. \cite{Wu21TFROM} employ a rank-based measure (NDCG) as the underlying utility for quantifying an individual's recommendation quality, whereas \cite{rastegarpanah2019fighting} use the mean squared error over a user's known ratings. \vspace{1.5mm}\\
        The variance of user/item deviation cost & Some works connect the notion of utility with the concept of \textit{cost}. For example, in~\cite{koutsopoulos2018efficient} state that to guarantee a minimum degree of item coverage, e.g., $d$-coverage, at least $d$ users must be recommended an item. The items in the recommendation list must be re-ranked in order to ensure an optimal ranking under such constraints. Individual fairness is defined by the cost of deviation from a nominal RS that does not account for item coverage and requires the incurred cost of deviation to be as evenly distributed across items or users as possible.\vspace{1.5mm}\leavevmode\\
        Based on Rawlsian fairness
        &
        Under Rawlsian principles \citep{rawls2001justice} of \textit{justice as fairness} and the \textit{difference principle} (where only inequalities that work to the advantage of the worst-off are permitted),
        the Max-min opportunity fairness metric \citep{Zhu21NewItems} accepts inequalities and aims to maximize the minimum utility of individuals or groups so that no subject is underserved by the model; for this, the average true positive rate of the $t\%$ worst-off items is computed, which are the $t\%$ items with the lowest true positive rates among all cold start items during testing.
       \leavevmode\\

    \midrule
    \multicolumn{2}{l}{\bf Group recommendation scenario}\leavevmode\\
    \midrule
        Aggregating effectiveness metrics on a group basis &
        Some authors aggregate metrics like NDCG or recall according to the users who belong to the same group. For these aggregations, the minimal value in a group or the ratio between minimal and maximal values have been used to quantify the gap between the least and highest utilities of group members in order to achieve social welfare \citep{Malecek2021Group,kaya2020ensuring}. \vspace{1.5mm}\leavevmode\\
        Other uses of effectiveness metrics &
        In group recommendation settings, where the recommendations for all the users in a group are combined into the same ranking, effectiveness metrics are used as surrogates of fairness to account for how many users are positively impacted by the recommendation. This is done in a way that higher values (more hits, or relevant recommendations for users) mean fairer recommendation lists. Such an approach was chosen in~\cite{Xiao20SocialActiv} with the average reciprocal hit rank, and in~\cite{kaya2020ensuring} with the zero-recall metric, which considers how many users received no relevant recommendations. Hence, lower values indicate fairer situations.
        \vspace{1.5mm} \leavevmode\\
        Satisfaction & Producing recommendations to a group should be fair when multiple iterations are allowed (sequential recommendation). In this context, the authors of~\citep{stratigi2020fair} propose several metrics to account for fairness: the overall satisfaction of a user (average of recommendation quality received by a user on each iteration), overall group satisfaction (average of overall user satisfaction across the group), and group disagreement (difference between maximum and minimum satisfaction values in a group).\leavevmode\\
\bottomrule
\end{longtable}
\color{black}
\normalsize

\paragraph{Discussion}
The main problem when using computational metrics in offline experiments, in general, is that it is often unclear to what extent these metrics translate to better systems in practice. In non-fairness research, this typically amounts to the question if higher prediction accuracy on past data will lead to more value for consumers or providers, e.g., in terms of user satisfaction or business-oriented key performance indicators, see~\citep{jannachjugovactmis2019}. In fairness research, the corresponding questions are if users would actually consider the recommendations fairer or if a fairness-aware algorithm would lead to the different behavior of the users. Unfortunately, research that involves humans is very rare. An example of a work that considers the effects of fair rankings can be found in~\cite{suhr2021does}, where
mixed effects were observed \review{in the context of job recommendation, accounting for gender biases and the impact of job context, candidate profiles, and employer inherent biases, revealing that fair algorithms are useful unless employers evidence strong gender preferences}.

Another potential issue of the \review{metrics used} is that they may be a strong over-simplification or too strong abstraction of the real problems. Consider the problem of recommending long-tail (less popular) items, which is in the focus of many research works. The metrics we found that measure how many long-tail items are recommended usually do not differentiate whether the recommended item is a ``good'' one or not, by using some form of quality assessment. As mentioned, some items may be unpopular just because of their poor quality.
Also, in many of these works, it is not clear what a desirable level of exposure of long-tail items would be. This is a problem that is particularly pronounced also for many works that measure fairness through the deviation of the recommendations from some target (desirable) distribution. In technical terms, adjusting the recommendations to be closer to some target distribution can be done with almost trivial and very efficient means like re-ranking. The true and important question, however, is how we know the target distribution in a given application context.

Generally, we also found a number of works where biased recommendations (e.g., towards popular items) were equated with unfairness. As discussed, this assumption may be too strong. In some of these papers, no deeper discussion is provided about why the biases lead to unfairness in a certain application context.
\review{The normative claims and underlying assumptions about how and when fairness is defined are missing, in parts leading to the impression that the concept of 'bias mitigation' instead of 'fairness' should have been used.}
\review{As noted earlier, a similar observation can be made for papers that assume that calibrating recommendations \textit{per se} leads to fairness. This can probably not be safely stated in general unless the normative claims are made explicit and fit the goals that are achieved by calibration.}

When considering recommendation quality metrics for groups, the assumption is either that different groups should have \review{equal recommendation quality (to treat them all alike) or that there is some justified inequality. The latter case may, for example, arise if some groups are assumed to receive better service, e.g., because they have paid for better service or when the inequality is dependent on the corpus size or the available relevant data~\citep{Kirnap21Estimation,amigo2022unifying}.} 

As argued above, in most applications of recommenders the recommendations will be better in terms of accuracy measures for active users than for less active users. Some papers in this survey consider this unfair, but this line of argumentation is not easy to follow.
\review{In fact, some researchers may argue that the correct mitigation strategy would be to fix the data or change the user interface to elicit more data. It would also be debatable which percentage of performance is acceptable to consider such a tradeoff (un)fair, as is the norm in the discussion around statistical parity.}
Certainly, there may be scenarios where there are particular protected attributes for which it may be desirable not to have largely varying accuracy levels across the groups. In many of the surveyed papers, no realistic use cases are however given.

\review{In terms of the different notions of fairness, traditionally either \textit{group fairness} or \textit{individual fairness} are studied to address consumer effectiveness and producer exposure. However, recent research also addresses situations involving \textit{mixed} individual and group fairness, such as group item exposure fairness and user-individual effectiveness fairness, see for example~\citep{Wu21TFROM,rastegarpanah2019fighting}. In such studies, it is often assumed that when provider exposure is addressed, the quality of the recommendations may diminish. The authors thus define individual unfairness as disparities in \textit{user losses} and demand that the decline in recommendation quality be dispersed equitably across all users. As previously stated, the notion of a \textit{trade-off} between the fairness evaluation objectives and overall system accuracy is prevalent in fairness research, and these demonstrate the need for additional research on multi-sided recommendation fairness.} 

Finally, looking at individual fairness in group recommendation scenarios, a multitude of aggregation strategies were proposed over the years such as Least Misery or Borda Count \citep{Masthoff2011GroupRS}. The literature on group recommender systems---which is now revived under the term fairness---however, does not provide a clear conclusion regarding which aggregation metric should be used in a given application.
\review{It should be noted that Arrow's impossibility theorem (from \emph{Social Choice Theory}) supports the conclusion that no aggregation strategy will be universally ideal, hence leading again to a potential reason for unfairness in a group.}
Also in this area researchers, \review{may} have been stuck in an abstraction trap~\citep{selbst2019,jannach2021mcnamara} \review{as we have pointed out several oversimplification instances in fairness research,} and more (multi-disciplinary) research seems required to understand group recommendation processes, see~\citep{DelicNNR18} for an observational study in the tourism domain.

\begin{figure*}[tb]
     \centering
     \includegraphics[width=0.6\linewidth]{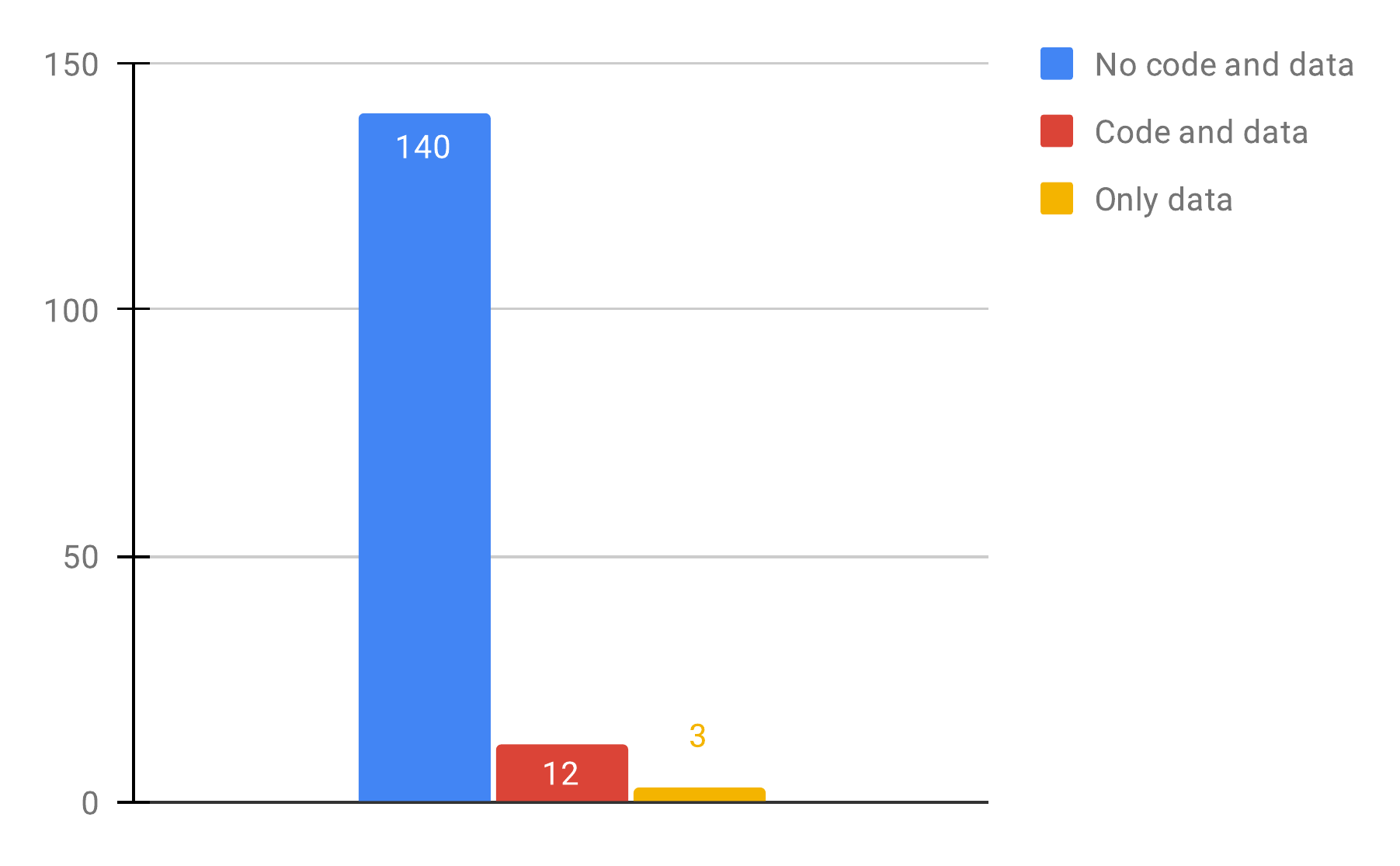}
     \caption{Level of Reproducibility (Shared Artifacts).}
     \label{fig:reproducibility}
 \end{figure*}

\paragraph{Reproducibility} The lack of reproducibility can be a major barrier to achieving progress in AI \citep{DBLP:conf/aaai/GundersenK18}, and recent studies indicate that limited reproducibility is a substantial issue also in recommender systems research \citep{cremonesi2021aimag,DBLP:journals/umuai/BelloginS21}.
Figure~\ref{fig:reproducibility} shows how many of the studied \emph{technical} papers and artifacts were shared to ensure the reproducibility of the reported experiments.\footnote{\review{The level of reproducibility of research work can be assessed in multiple dimensions, see \citep{DBLP:conf/aaai/GundersenK18}. In the context of our work, we limit ourselves to the analysis of certain central artifacts that are publicly shared.}}
While the level of reproducibility seems to be higher than in general AI \citep{DBLP:conf/aaai/GundersenK18}, still for the large majority of the considered works authors did not share any code or data.

\subsection{Landscape Overview}\label{ss:landscape}

\begin{figure*}[tb]
     \centering
     \includegraphics[width=\linewidth]{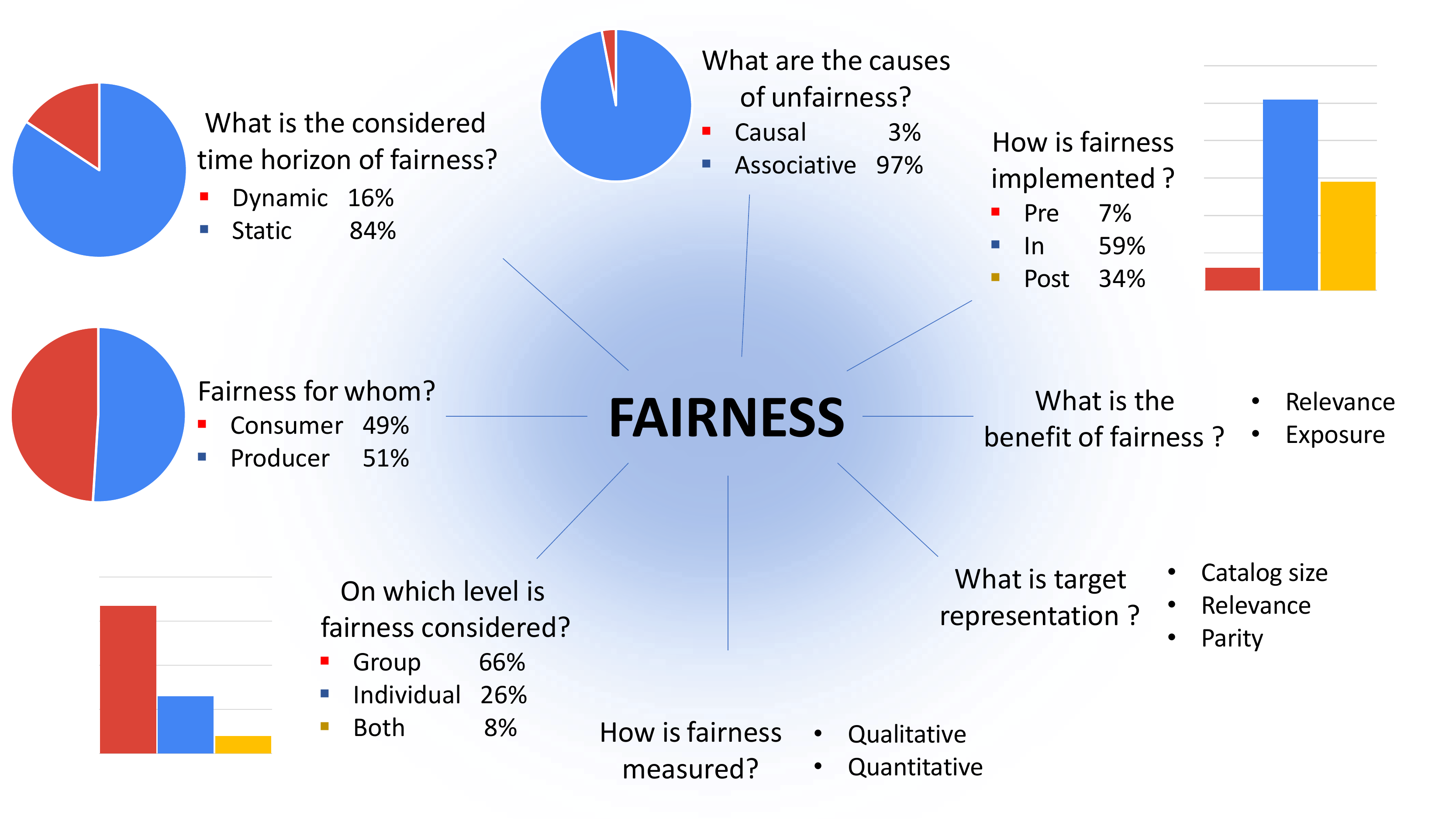}
     \caption{Taxonomy and Landscape.}
     \label{fig:statistics-overview}
 \end{figure*}

Fairness is a multi-faceted subject. In order to provide an encompassing understanding of different fairness dimensions, we have developed
\review{a taxonomy that takes different perspectives, as explained in Section \ref{ss:prev}, which allows us to describe the}
landscape of fairness research in recommender systems, as shown in Figure \ref{fig:statistics-overview}. The landscape's main aspects can be summarized based on the following questions.

\begin{itemize}
\item \textbf{How is fairness implemented?} Depending on which step of the recommendation pipeline we change, fairness-enhancing systems can be divided into are pre-, in- and post-processing techniques. Here we also note that the main patterns are in- and post-processing (typically re-ranking), probably due to the advantage of an easier applicability to existing systems.

\item \textbf{What is the target representation?} The \textit{target representation} is defined as the ideal representation (i.e., proportion or distribution of exposure) \citep{Kirnap21Estimation}. In other works, this is also referred to as \textit{target distribution} (of benefits such as exposure or relevance). Even though this aspect has not been specifically analyzed in the previously presented figures, we have identified three main target representations against which most fairness metrics compare: catalog size, relevance, and parity. These representations match those introduced in \cite{Kirnap21Estimation}, where authors state that the choice of the representation target depends on the application domain. Among these, the most common interpretation is that items should be recommended equally for each group, hence, using a parity-based representation target. However, there are also other aspects and fairness notions that do not use this assumption, as discussed in Section~\ref{ss:notions}.

\item \textbf{What is the benefit of fairness?}
As in the previous case, for the sake of conciseness, we have not considered this dimension in this detailed analysis, but it is worth mentioning that fairness definitions can be categorized depending on whether its main benefit is based on exposure (by assessing if items are exposed in a uniform or fair way) or relevance (with the additional constraint on the exposure that it must be effective, that is, it should match the user preferences). In principle, any information seeking system (such as search engines or recommender systems) should aim for relevance-based benefits. However, considering the difficulty of these tasks, by measuring and achieving a situation with fair exposure, the subsequent measurements on the system would already be impacted and improved, from a fairness perspective and, hence, it is a reasonable goal to obtain.

\item \textbf{How is fairness measured?}
Fairness evaluation, as any other experimental research, can be performed through qualitative or quantitative methods. As discussed in Section~\ref{ss:meth}, qualitative approaches are currently almost never taken, and most of the analyses are done by quantitative approaches such as offline experiments or A/B tests.

\item \textbf{On which level is fairness considered?} Fairness can be defined on a group level or individual level, as discussed above. Today, group-level fairness is the prevalent option, most likely because  measuring (operationalizing) group fairness is easier than individual fairness. In other words, what it means for two individuals to be similar is task-sensitive and more difficult than segmenting users/items into groups based on a sensitive feature, as is often done in the examined literature of group fairness. This might also have social implications, as many major considerations of fairness in the literature, including gender equality, demographic equality, and others, are predicated on the concept of group fairness.
\review{This is connected with the so-called \textit{issue of intersectionality}, which we discuss in some more detail below.}
It is important to note that the primary limitation of group fairness is the decreasing reliability of sensitive attributes in recent years due to privacy concerns and firms' reluctance to share such information.

\item \textbf{Fairness for whom?} In many cases, the circumstance for making a recommendation is intrinsically multi-sided. As a result, any of the \textit{stakeholders} engaged, as well as the platform itself, may be affected by (un)fairness. Through our survey, we found that there is a balance in the literature
between consumer and provider viewpoints.
\review{In addition, more recent research in ML has begun to address the issue of \textit{intersectionality} in fairness by building statistical frameworks that account for bias within multiple protected groups, for
example, \dquotes{black women} instead of just \dquotes{black people} or \dquotes{women}~\citep{ghosh2021characterizing,morina2019auditing}. An interesting example is presented by~\cite{buolamwini2018gender} where the authors found that commercial facial image classification systems do not show the full distribution of mis-classifications when considering gender and skin type alone, and that darker-skinned women being the most misclassified group, with an accuracy drop of over 30\% compared to lighter-skinned men. This aspect has, to the best of our knowledge, been largely overlooked in the recommender fairness research; one exception is the study presented recently by ~\cite{shen:ipm23}, where such intersectionality between gender (male vs.~female) and skin color (black vs.~white) fairness was applied to  language model-driven conversational recommendation.}

\item \textbf{What is the considered time horizon of fairness?} Fairness can be pursued in a static way (or: \textit{one-shot}), or dynamically over time, taking into account shifts in the item catalog, user tastes, etc. However, practically we observe a prevalence of the former, with the latter including new trends like reinforcement learning-based approaches.

\item \textbf{What are the causes of unfairness?}
The dominant pattern of fairness-enhancing approaches seems to pursue a static, associative, group-level notion of fairness, inheriting from fair ML traditional research. Hence, papers considering relatively new approaches such as causal inference and long-term fairness are more rare. We can describe this as a research gap, i.e., there should be more research into the reasons of unfairness through the lens of causality and counterfactuals.

\end{itemize}

\section{Discussion}\label{sec:discussion}
\paragraph{Summary of Main Observations}
Due to today's broad and increasing use of AI in practical applications, questions relating to the potential harms of AI-powered systems have received more and more attention in recent years, both in academic research, the tech industry, and within political organizations. Fairness is often considered a central component of what is sometimes called \emph{responsible} AI. These developments can also be seen in the area of recommender systems, where we observed a strong increase in terms of publications on fairness since the mid-2010s, cf.~Figure~\ref{fig:historical-development}.

Looking closer at the research contributions from the field of computer science, we observe that the large majority of works aim to provide technical solutions, and that the technical contributions are predominantly fairness-aware algorithms (cf.~Figure~\ref{fig:technical-vs-conceptual} and Figure~\ref{fig:technical-focus}).
In contrast, only comparably limited research activity seems to take place on topics that go beyond \review{the algorithmic perspective, such as user interfaces and human-in-the-loop approaches, or even beyond computer science (that is applied to AI in general, and recommender systems in particular), such as psychology, economics, or social sciences.}
While algorithmic research is certainly important, focusing almost exclusively on improving algorithms in terms of optimizing an abstract computational fairness metric may be too limited. Ultimately, however, our goal should rather be to design ``\emph{algorithmic systems that support human values}''~\citep{fairness2018tutorial} and avoid potential abstraction traps, similar as in the general area of fair ML.

On the positive side, we find that researchers in fair RS are addressing various notions of fairness (cf.~Figures~\ref{fig:individual-vs-group-fairness}
to \ref{fig:associative-vs-causal-fairness}), e.g., they deal with questions both of individual fairness and of group fairness.
In addition, the community has expanded the scope of fairness considerations beyond
\review{its impact on people}
and has developed various approaches to deal with fairness towards items and providers.
This is different from many other traditional application areas of fair ML, e.g., credit default prediction, where
\review{people}
are usually the main focus of research\review{, even though these concepts of item fairness are ultimately always related to people (or organizations) in the end, because the item providers are the ones impacted when their items are not recommended}.

Looking at the considered application domains and datasets, we observe that various domains are addressed. However, the large majority of technical papers report experiments with datasets from the media domain (videos and music), cf.~Figure~\ref{fig:application-domains}. Specifically, some of the MovieLens datasets are frequently used either as a concrete use case or as a way to at least provide reproducible results, given that the set of fairness aspects that can be reasonably studied with such datasets seems limited. All in all, there seems to be a certain lack of real-world datasets for real-world fairness problems, which is why researchers frequently also rely on synthetic data or on protected groups that are artificially introduced into a given recommendation dataset.

In terms of the research methodology, offline experiments using the described datasets are the method of choice for most researchers, cf.~Figure~\ref{fig:experiment-types}. Only very few works rely on studies that have the human in the loop, \emph{which points to a major research gap in fair recommender systems}. In the context of these offline evaluations, a rich variety of evaluation approaches and computational metrics are used. The way the research problems are operationalized however often appears to be an oversimplification of the underlying problem. In many research works, for example, (popularity) biases are equated with unfairness, which we believe is not necessarily the case in general. Some of the surveyed works also seem to ``re-brand'' existing research on beyond-accuracy quality aspects of recommendations---e.g., on diversity or calibration---as fairness research,
\review{sometimes missing a clear and detailed discussion of the underlying normative claims that are addressed.}
Finally, in almost all works some ``gold standard'' for fair recommendations is assumed to be given, e.g., in the form of a target distribution regarding item exposures. With the goal of providing generic algorithmic solutions, little or no guidance is however usually provided on how to decide or determine this gold standard for a given use case. While general-purpose solutions are certainly desirable,
the danger of being stuck in \review{an} abstraction trap with limited practical impact increases \citep{selbst2019,jannach2021mcnamara}.

\paragraph{Future Directions}
Our analysis of the current research landscape points to a number of further research gaps. Considering the type of contributions and the different notions of fairness, we find that today's research efforts are not balanced. Most published works are algorithmic contributions and use offline evaluations with a variety of proxy metrics to assess fairness. \review{Less discussion is provided regarding how different level content used in mainstream recommender systems (e.g., user-generated, expert-generated content, and audio) \citep{moscati2022music4all,deldjoo2021content} are susceptible to the promotion of certain types of biases and unfairness, e.g., audio content could suffer more from an accuracy standpoint but could promote the recommendation of long-term items more effectively.} Moreover, these offline evaluations are based on one particular point in time. As such, these evaluations do not consider longitudinal dynamics that may emerge (a) when the fairness goals change over time or (b) when an algorithm's output changes over time, e.g., when a fairness intervention gradually improves the recommendations. This limitation of static offline evaluations also becomes more acknowledged in the general recommender systems literature. Simulation approaches are recently often considered as one promising approach to model such longitudinal dynamics \citep{ghanem2022balancing,rohde2018recogym,mladenov2021recsimng,longitudinalimpact2021}.
Causal models, in contrast to associative ones, also received very limited research attention so far.

Through our survey, we furthermore identified a number of promising research problems for which only few works exist so far:

\begin{itemize}
   \item \textbf{Challenge 1:} \emph{\review{Achieving realistic and useful definitions for fairness.}}
   As discussed before, there are several definitions for fairness, not only in the RS literature but in ML and AI in general \review{\citep{DBLP:journals/fdata/Olteanu00K19}}.
   This provokes incompatibility between some of these definitions and potential disagreement, where one metric may conclude that a recommender system is fair and another the opposite, even from a mathematical point of view \citep{DBLP:journals/bigdata/Chouldechova17}.
   As a consequence, it is not easy to find a proper balance between different notions of fairness and the performance of the recommendation models. An example of a relevant proposal can be found in \cite{DBLP:conf/pakdd/LiuLTLCH20}, where the authors employ metrics that capture the cumulative reward in a way that combines accuracy and fairness while aiming to improve both.
   \review{This is a rich area of investigation, open to novel definitions and approaches about how to leverage this tradeoff and whether one dimension should weight more than the other \citep{DBLP:journals/cacm/FriedlerSV21,DBLP:journals/bigdata/Chouldechova17,DBLP:conf/innovations/KleinbergMR17}.}

   However, this is not the only problem we have identified in our literature review. As stated in Section~\ref{ss:meth}, the seldom use of user studies and field tests make it
   \review{very difficult}
   to incorporate user perception \citep{chiir:2023} into our understanding of what should be defined as a fair recommendation.
   In fact, some works propose to move from notions of equality to those of equity and independence~\citep{amigo2022unifying}, but even these general definitions that may work at a societal level, may not necessarily make sense depending on the domain or the user needs.

   \item \textbf{Challenge 2:} \emph{Building on  appropriate data to assess fairness.}
    As discussed in Section~\ref{ss:domains}, some datasets used in the literature do not contain sensitive attributes at all. This problem has been addressed in different ways, none of them perfect but fruitful towards the goal of mimicking the evaluation of recommender systems in realistic scenarios.
    A first possibility is to perform data augmentation, where the main idea is, without changing the underlying data and algorithm, to be able to remove biases from the data to provide higher-quality information to the algorithms \citep{rastegarpanah2019fighting}.
    Another, more popular, possibility is to use of simulation instead of real-world datasets. Various recent papers use simulation, sampling techniques (see e.g., the work by \cite{deldjoo2021explaining} investigating the impact of data characteristics), and synthetic data to evaluate fairness in search scenarios ~\citep{DBLP:conf/kdd/GeyikAK19}. This may require more advanced techniques in the evaluation step, such as counterfactual evaluation, in order to properly interpret the data coming from A/B logged interactions once interventions have been performed through a recommendation algorithm, for example, by focusing on improving item exposure~\citep{Mehrotra18Towards}.

   \item \textbf{Challenge 3:} \emph{Understanding fairness in reciprocal settings.}
    Maintaining the utility of stakeholders in reciprocal settings is a new notion of fairness \citep{DBLP:journals/kbs/XiaYXL19}, even though reciprocal recommender systems have been studied (although not as frequently as other systems) in the past and remain at the core of social network and matching platforms, \review{see \citep{DBLP:reference/sp/KoprinskaY15} for a survey on people-to-people recommender systems}.
    In     \review{the former} work,
    \review{Xia et al.} define fairness as an equilibrium between parties where there are 'buyers' and 'sellers' and each seller has the same value or 'price'; hence, in their notion of \dquotes{Walrasian Equilibrium} they are treated fairly by considering at the same time (a) the disparity of service, (b) the similarity of mutual preference, and (c) the equilibrium of demand and supply\review{, that is, by balancing the demand of buyers and the supply of sellers}.

    By considering the importance of this type of systems, being able to operationalize a reasonable definition for this context is foreseen as a major challenge to tackle in the future.
    \review{In fact, going beyond these notions of equilibrium for reciprocal settings, such as cooperative behaviors and non-zero sum games, would require digging further into game theory and related areas, which would be potential avenues for future research.}

   \item \textbf{Challenge 4: } \emph{Fairness auditing.}
    As stated in \cite{DBLP:journals/computer/KoshiyamaKT22}, algorithm auditing is the research and practice of assessing, mitigating, and assuring an algorithm’s legality, ethics, and safety.
    In that work, the authors consider bias and discrimination as one of the main verticals of algorithm auditing.
    Hence, auditing recommender systems should become a priority in the near future, and the fairness dimension is, by definition, one of the most important aspects to be considered in that process.
    As an example, we want to highlight that the authors from \cite{DBLP:conf/bias/KrafftHZ20} aimed at auditing decision making systems, but faced important issues since their agents were banned from the platform that was meant to be analyzed (Facebook NewsFeed). Hence, there are technical difficulties that may make this challenge even harder to achieve, despite its importance in legal and ethical dimensions.
    \review{Because of this, we argue that, in order to be practical and potentially address this challenge, such requirements should be enforced from higher levels or even policies, otherwise companies may not embrace this type of accountability.}
\end{itemize}

Finally, one main fundamental problem of current research on fair recommender systems is that it is not entirely clear yet how impactful it is in practice. Algorithmic research is too often based on a very abstract and probably overly simplistic operationalization of the research problem, using computational metrics for which it is not clear if they are good proxies for fairness in a particular problem setting. In such a research approach, fundamental questions of what is a fair recommendation in a given situation are not discussed. Correspondingly, the choice of application domains sometimes seems arbitrary (based on dataset availability), and the fairness challenges often appear almost artificial. Moreover, connections to existing works and theories developed in the social sciences are rarely established in the published literature, and fairness is often simply treated as an algorithmic problem, e.g., to make recommendations that match a pre-defined target distribution. In some ways, current research shares challenges with many works in the area of Explainable AI, where many insights from social sciences exist, and where it is often neglected that explainable AI, like a recommendation, to a large extent is a problem of human-computer interaction \citep{MILLER20191}. As a consequence, much more fundamental research on fairness, its definition in a given problem setting, and its perception by the involved stakeholders is needed.
This, in turn, requires a multidisciplinary approach, involving not only researchers from different areas of computer sciences, but also including subject-matter experts from real-world problem settings and
scholars from fields outside computer science,
\review{such as psychology and social science.}

\begin{acknowledgements}
The authors thank the reviewers for their thoughtful comments and suggestions.
\end{acknowledgements}

\bibliographystyle{named}
\bibliography{ijcai22}

\end{document}